\documentclass[twocolumn,apj,numberedappendix,twocolappendix]{openjournal}

\usepackage{amsfonts, amssymb,amsmath,bm}

\usepackage{xspace} 
\usepackage{booktabs}
\usepackage{graphicx}
\usepackage{color}
\usepackage{ulem}
\usepackage{multirow}
\definecolor{darkblue}{rgb}{0.0, 0.0, 0.55}
\definecolor{darkred}{rgb}{0.55, 0.0, 0.0}
\definecolor{darkviolet}{rgb}{0.2,0,0.2}
\definecolor{ForestGreen}{rgb}{0.13, 0.55, 0.13}
\definecolor{airforceblue}{rgb}{0.36, 0.54, 0.66}
\definecolor{orange}{rgb}{1.0, 0.5, 0.0}
\definecolor{alizarin}{rgb}{0.82, 0.1, 0.26}
\definecolor{brilliantlavender}{rgb}{0.96, 0.73, 1.0}

\definecolor{color1}{rgb}{0.3, 0.7, 0}

\usepackage{hyperref}
\hypersetup{
    colorlinks=true,
    linkcolor=darkblue,
    citecolor= blue,   
    filecolor=blue,      
    urlcolor=magenta,
    pdftitle={Overleaf Example},
    pdfpagemode=FullScreen,
}

\newcommand{\de}{\partial}

\newcommand{\euclid}{\textit{Euclid}\xspace}
\newcommand{\halofit}{\textsc{Halofit}\xspace}
\newcommand{\hmcode}{\textsc{HMcode}}

\newcommand{\cosmomentum}{\textsc{CosMomentum}\xspace}
\newcommand{\ldt}{\textsc{L2DT}\xspace}
\newcommand{\pyccl}{\texttt{pyccl }}
\newcommand{\treecorr}{\texttt{TreeCorr}}

\newcommand{\cyl}{\text{cyl}}

\newcommand{\fid}{\text{fid}}
\newcommand{\noisy}{\text{noisy}}
\newcommand{\noisefree}{\text{noise\-free}}

\begin{document}
	
\title{Unleashing cosmic shear information with the tomographic weak lensing PDF} \author[0000-0002-2324-7335]{Lina Castiblanco$^{1,\star}$}
\author[0000-0001-7831-1579]{Cora Uhlemann$^{1,2}$}
\author[0000-0002-4864-1240]{Joachim Harnois-D\'eraps$^1$}
\author[0000-0003-1060-3959]{Alexandre Barthelemy$^3$}
\thanks{$^\star$\href{mailto:lina.castiblanco-tolosa@newcastle.ac.uk}{lina.castiblanco-tolosa@newcastle.ac.uk}}

\affiliation{$^{1}$ School of Mathematics, Statistics and Physics, Newcastle University, Herschel Building, NE1 7RU Newcastle-upon-Tyne, U.K.}
\affiliation{$^2$ Fakultät für Physik, Universität Bielefeld, Postfach 100131, 33501 Bielefeld, Germany}
\affiliation{$^3$ Universit\"{a}ts-Sternwarte, Fakult\"{a}t f\"{u}r Physik, Ludwig-Maximilians-Universit\"{a}t M\"{u}nchen,\\Scheinerstra{\ss}e 1, 81679 M\"{u}nchen, Germany}

\begin{abstract}
    In this work, we demonstrate the constraining power of the tomographic weak lensing convergence PDF for StageIV-like source galaxy redshift bins and shape noise. We focus on scales of $10$ to $20$ arcmin in the mildly nonlinear regime, where the convergence PDF and its changes with cosmological parameters can be predicted theoretically. We model the impact of reconstructing the convergence from the shear field using the well-known Kaiser-Squires formalism. 
    We cross-validate the predicted and the measured convergence PDF derived from convergence maps reconstructed using simulated shear catalogues. Employing a Fisher forecast, we determine the constraining power for $(\Omega_{m},S_{8},w_{0})$. We find that adding a 5-bin tomography improves the $\kappa-$PDF constraints by a factor of $\{3.8,1.3,1.6\}$ for $(\Omega_{m}, S_{8},w_{0})$ respectively. Additionally, we perform a joint analysis with the shear two-point correlation functions, finding an enhancement of around a factor of $1.5$ on all parameters with respect to the two-point statistics alone. These improved constraints come from disentangling $\Omega_{\rm m}$ from $w_0$ by extracting non-Gaussian information, in particular, including the PDF skewness at different redshift bins. We also study the effect of varying the number of parameters to forecast, in particular we add $h$, finding that the convergence PDF maintains its constraining power while the precision from two-point correlations degrades by a factor of $\{1.7,1.4,1.8\}$  for $\{\Omega_{\rm m},S_8,w_0\}$, respectively.
\end{abstract}

\section{Introduction}
Stage IV galaxy surveys such as \euclid  \citep{EUCLID:2011zbd} or the \textit{Vera C. Rubin} Observatory Legacy Survey of Space and Time \citep[][LSST]{Ivezić_2019}, SPHEREx \citep{SPHEREx:2014bgr}, Dark Energy Spectroscopic Instrument (DESI, \cite{DESI:2016fyo}), The \textit{Nancy Grace Roman} Space Telescope \citep{Spergel:2015sza}, promise unprecedented insight into the physics and evolution of the universe.
While the good redshift resolution of spectroscopic surveys provides access to 3D galaxy clustering, photometric surveys can map a larger number of galaxies including their shapes to probe the overall matter distribution. With tomographic techniques, photometric surveys can also track the growth of structure and thus probe signatures of dark energy. These next-generation of telescopes will provide a large volume and quality data that will significantly improve our understanding of dark energy, dark matter and the initial conditions of the universe. 

Gravitational weak lensing is one of the most powerful and accurate probes of the large-scale structure of the universe \citep{Kilbinger:2014cea}. Weak gravitational lensing traces the projected distribution of dark matter through the gravitational effect on the light emitted by distant source galaxies, which translates into cosmic shear \citep[the deflection in the shape and size of the observed galaxies, see][]{Kaiser:1992ps, Bartelmann:1999wg}. Standard weak lensing provides a 2D map of the second derivatives of the projected gravitational potential, from which it is not possible to fully recover the time evolution of the density field. Weak lensing tomography recovers some of the lost information by dividing the galaxy distribution into different redshift bins \citep[see e.g.][]{Munshi:2011bf,Jee:2015jta,Martinet:2020omp}. The weak lensing signal at different source redshifts is sensitive to the nonlinear growth of structure influenced by dark energy \citep{Sipp:2020pfx} and massive neutrinos, which leave a strong signal on small scales \citep{Liu2019}. Therefore, cosmic shear observations provide great precision in constraining  the parameters that describe the content and evolution of the universe. 

Analysing upcoming high-quality cosmological data to its fullest potential represents a significant challenge, especially when aiming to capture the non-Gaussian late time information stored in small scales. Extracting the weak lensing signal relies on statistical measurements of correlations between galaxy shapes. Traditionally, the shear two-point correlation function (2PCF) or its Fourier transform, the lensing power spectrum, are employed for this \citep{Joachimi:2007xd,Kilbinger:2012qz,Hamana:2019etx,KiDS:2020suj,DES:2021bvc,DES:2022qpf}. Both provide valuable information about the spatial distribution and clustering properties of matter in the universe. However, the non-Gaussian information encoded in the small scales are not captured by the 2PCF. To unlock the full potential of Stage IV surveys and improve the  constraints on cosmological parameters higher order statistics (HOS) are crucial. The HOS capture additional information providing a significant enhancement on the constraints when combined with the 2PCF \citep{Euclid_overview_2024}. Some examples of HOS applied to lensing data are: Higher-order moments of the convergence or aperture mass \citep{VanWaerbeke:2013eya, Porth:2021lyg,DES:2023qwe}, shear/convergence peak counts \citep{Marian:2008fd,Dietrich:2009jq,DES:2016jfa,Martinet:2017rqp,Harnois-Deraps:2020pvj,DES:2021epj,Marques:2023bnr},  Minkowski functionals \citep{Ajani:2022ifk, Vicinanza:2019fzo, Petri:2015ura, Kratochvil:2011eh}, Betti numbers \citep{Parroni:2020xtj}, persistent homology \citep{Heydenreich:2020hrr, Wilding:2020oza}, scattering transform coefficients \citep{ChengChengSiHao:2021hja, DES:2023qwe} and the one-point probability distribution function \citep{Barthelemy:2019ciu, Boyle:2020bqn,Barthelemy:2023mer}. The constraining power of this set of HOS has been recently investigated  for the \euclid survey in \cite{Euclid:2023uha}. The authors performed a non-tomographic Fisher forecast for the matter density parameter $\Omega_{\rm m}$ and the amplitude of fluctuations $\sigma_8$ and demonstrated the potential of each of the HOS over the 2PCF due to their  ability to extract non-Gaussian, small scale information. Most of the HOS improve over the 2PCF constraints with slightly different degeneracy directions and partial redundancy between different HOS.

In this work, we focus on the weak lensing convergence one-point probability distribution function (PDF) which unlike many HOS can be predicted theoretically. The weak lensing convergence, $\kappa$, offers a direct way to trace the projected matter distribution along the line-of-sight  unaffected by how galaxies trace the underlying matter distribution.  The kappa-PDF offers valuable information beyond two-point analyses, potentially improving constraints on cosmological parameters. It can be easily measured from simulated and observed $\kappa$-maps and can be modelled with high accuracy on mildly non-linear scales. The theoretical model of the PDF based on the large deviation theory (LDT) has been applied on 3D density fields to describe dark matter and halo densities and showed strong potential to constrain different $\Lambda$CDM parameters \citep{Bernardeau:2015khs,Uhlemann:2015npz, Codis:2016gyt, Uhlemann:2017hgi,Uhlemann:2019gni} as well as the total neutrino mass $M_\nu$ \citep{Uhlemann:2019gni}, the primordial skewness $f_{NL}$ \citep{Uhlemann:2019gni, Friedrich:2019byw} and dark energy $(w_0,w_a)$ and modified gravity parameters \citep{Cataneo:2021xlx,Gough:2021hlr}.  

The LDT has been also applied for projected fields in 2D for different probes, including  the galaxy weak lensing convergence \citep{Barthelemy:2019ciu}, the CMB lensing convergence \citep{Barthelemy:2020igw}, the aperture mass field  \cite{Barthelemy:2020yva}, the projected galaxy density \citep{Uhlemann:2017hgi,Friedrich:2021xff} and density-split statistics \citep{DES:2017hhj,Gruen_2018, Burger:2022lwh}. The key physics component over which the  LDT is built is the spherical \citep{Valageas:2001zr,Bernardeau:2013dua} and cylindrical \citep{Bernardeau:2000et} gravitational collapse for 3D and 2D fields, respectively. \cite{Boyle:2022msq} validated the theoretical framework presented in \citep{Barthelemy:2019ciu}, extended it to include massive neutrinos and performed a non-tomographic Fisher analysis. This study finds that focusing on the bulk of the PDF achieves good constraining power surpassing the 2PCF while maintaining a Gaussian likelihood enabling Fisher forecasts. LDT  provides a direct theoretical model for all non-Gaussian contributions to the cumulant generating function (CGF) from which the PDF is obtained through an inverse Laplace transformation. \cite{Boyle:2022msq} take the first steps into studying the cosmological information contained in the weak lensing convergence CGF, showing it is an alternative statistics to access the non-Gaussian information at mildly non-linear scales.

The weak lensing convergence is not a direct observable, it has to be reconstructed from the shear field using mass mapping techniques such as the Kaiser-Squires (KS) inversion method \citep{Kaiser:1991qi}. This method uses the relationship between the spin-2 shear and the spin-0 convergence which is only explicit in harmonic (Fourier or spherical harmonics) space. However, when applied to real survey data with missing information (due to masks), the KS method presents some limitations, as it assumes a periodic and continuous space. When applied to finite and masked regions, it generates spurious frequencies at the borders, which leads to a loss of power on the reconstructed field as well as the loss of fundamental properties of the lensing fields with the creation of non-physical B-modes. 

This paper investigates the power of tomographic weak lensing convergence PDF for constraining cosmological parameters. We use the method proposed and validated in \cite{Barthelemy:2023mer} that follows a novel approach by incorporating the impact of KS-reconstruction directly into the modelling of the $\kappa-$PDF. 
We predict the $\kappa-$PDF for a StageIV-like source galaxy distribution and include shape noise. We employ equipopulated redshift bins, following the optimisation study performed in \cite{Sipp:2020pfx}. We propose  an optimised  tomographic strategy that matches the number of cross-correlation bins while combining consecutive bins to suppress shape noise. We show the accuracy of the theoretical predictions by validating them with the maps reconstructed from the SLICS shear catalogues \citep{Harnois-Deraps:2018bcv}. Using a Fisher forecast, we assess the constraining power of the tomographic $\kappa-$PDF for the set of parameters, $(\Omega_{\rm m},S_8=\sigma_8\sqrt{\Omega_{\rm m}/0.3},w_0)$ finding a large improvement compared with non-tomography analysis. Following the same procedure with the shear 2PCF we demonstrate the superiority and complementarity  of the $\kappa-$PDF over the 2PCF in constraining these set of parameters. In addition, we investigate the effect of varying the number of parameters to forecast. We find that the $\gamma$-2PCF loses constraining power when forecasting more than two parameters, while the kappa-PDF maintains its power, providing better constraints. 

Alternative theory and simulation methods have already shown the strong potential in the tomography $\kappa-$PDF to obtain valuable cosmological information. \cite{Thiele:2020rig} uses an analytical description based on a halo-model approach, which is accurate at small scales but not at mildly non-linear scales. A forecast for the neutrino masses $M_\nu$, base on simulations, is performed in \cite{Liu2019}. Their method is able to probe the internal structure of halos by accessing the one-halo regime. \cite{Giblin:2022ucn} uses a method based on Gaussian process regression to perform a full cosmological parameter estimation analysis using simulated lensing PDF measurements. Additionally, simulation-based methods combined with real data have also been applied to infer the value of $S_8$ \citep{thiele2023hsc} or to study the effect of baryons \citep{Grandon:2024pek}. This method allows including highly non-linear scales. Our goal is to demonstrate the accuracy and precision of our method, which is valid for a larger range of scales compared to other theoretical methods and less computationally expensive than simulation-based methods. 

This work is structured as follows: in section \ref{sec:theory} we present the theoretical framework of the summary statistics we use, i.e. the $\kappa-$PDF and the $\gamma-$2PCF. In  section \ref{sec:theory_validation} we describe the SLICS simulations and use them to validate our theoretical predictions. In section \ref{sec:Fisher} we describe the cosmology  behaviour of $\kappa-$PDF derivatives and the structure of the data covariance and show our results on the Fisher forecast. We conclude and provide an outlook on future work in section \ref{sec:conclusions}. We provide additional test in the Appendix section, in particular we include and study the effect of intrinsic alignments on the predicted $\kappa-$PDF in Appendix \ref{app:IA}.

\section{Theoretical framework for weak lensing statistics} \label{sec:theory}

Weak gravitational lensing by large scale structures gives rise to a phenomenon known as \textit{cosmic shear}, which manifests as a distortion in the apparent shapes of observed galaxy images. The deflection on the galaxy images is quantified in terms of three quantities: the shear components $(\gamma_1,\gamma_2)$ and the convergence $\kappa$, all three quantities are sourced by the projected lensing gravitational potential $\psi(\chi\bm{\theta},\chi)$\footnote{The projected gravitational potential along the line-of-sight is defined as $$\psi(\bm{\theta},\chi) = \frac{2}{c^2} \int^{\chi}_0 d\chi'\frac{f_{K}(\chi-\chi')}{f_K(\chi)f_K(\chi')}\Phi(f_K(\chi')\bm{\theta},\chi')\,,$$ where $\Phi$ is the Newtonian gravitational potential, $f_K(\chi)$ is the comoving angular distance. We consider a flat universe, therefore $\chi(z)$ is equivalent to the comoving angular distance $f_K(\chi)$.}. In Born's approximation these three quantities are given by \cite{Kilbinger:2014cea} 
\begin{equation}
    \gamma_1 = \frac{1}{2}(\de_1\de_1-\de_2\de_2)\psi\, \quad \gamma_2 = \de_1\de_2 \psi\, \quad \kappa =\nabla^2 \psi \,, \label{eq:shear_equations}
\end{equation}
with $(\de_1,\de_2)$ the partial derivatives with respect to $\bm{\theta} =  (\theta_1,\theta_2)$, the angular coordinates of the orthogonal planes to the light-of-sight $\bm{\hat{n}}$.

The weak lensing convergence $\kappa$ is defined as the line-of-sight projection of the matter density distribution between the observer and the galaxy source as 
\begin{figure}
    \centering
    \includegraphics[scale=0.45]{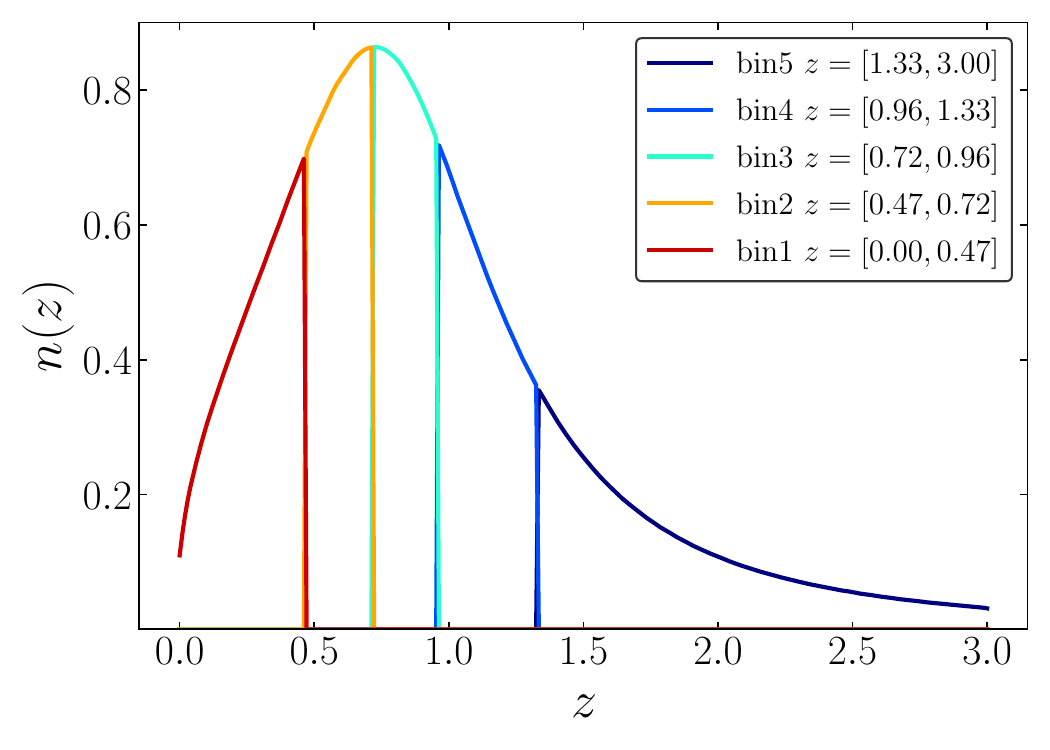}
    \includegraphics[scale=0.35]{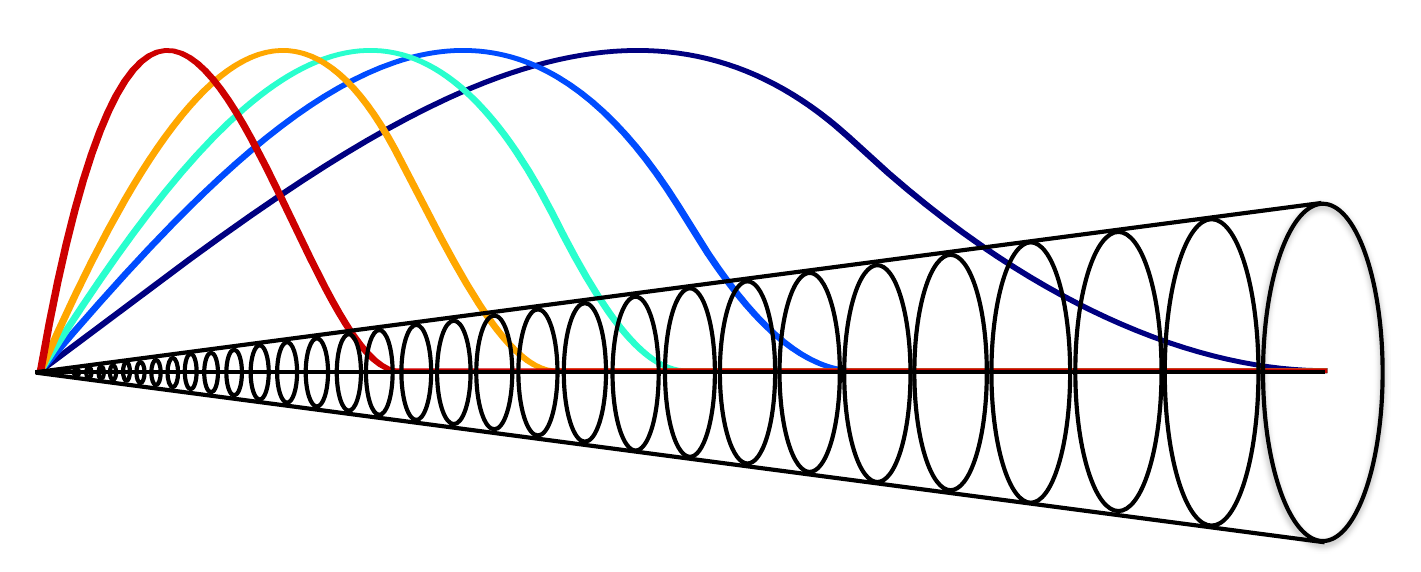 } 
    \caption{\textbf{Upper:} StageIV-like source redshift distribution probing up to source redshifts of $z_s\sim 3$. The redshift distribution is sliced into 5 equipopulated redshift bins. Ranges for each redshift bin are presented in the legend. \textbf{Lower:} Graphical representation of the convergence field in a fixed aperture as a superposition of independent layers of the 3D matter density along the line-of-sight. The contributions of the different slices are weighted according to the weak lensing kernel~\eqref{eq:weight_function}.}
    \label{fig:n_of_z}
\end{figure}
\begin{equation}
    \label{eq:kappa}
    \kappa(\bm{\theta}) =  \int_{0}^{\infty} \,d\chi \, w(\chi(z))\delta(\chi(z)\bm{\theta},\chi(z)) \,
\end{equation}
where $\chi(z)$ is the comoving distance to redshift $z$. For a population of galaxy sources the convergence is also weighted by its redshift distribution $n(z)$, such that the weak lensing projection weight function $w(\chi(z))$ is  given by 
\begin{equation}
    \label{eq:weight_function}
    w(\chi) = \frac{3\Omega_{\rm m} H_0^2}{2 c^2}\int_{z}^{\infty} \!\!\!\! dz' \frac{\left[\chi(z')-\chi(z)\right]\chi(z)}{\chi(z')}(1+z)n(z')\,
\end{equation}
with $H_0$ the Hubble parameter, $\Omega_{\rm m}$ matter density fraction and $c$ the speed of light. For the tomographic analysis, the redshift distribution of source galaxies $n(z_s)$ is divided into different tomographic bins, as shown in the upper panel of figure \ref{fig:n_of_z}. The weak lensing convergence in a fixed aperture can be conceptualised as the superposition of independent layers of the 3D matter density field in a cone as illustrated in the lower panel of  figure \ref{fig:n_of_z}. For correlators of the convergence this corresponds to the so-called Limber approximation.

The convergence quantifies changes in apparent size of the image (and the flux) while the shear components measure the change in the shape. Weak lensing affects both the shape and size of the observed galaxy images by a few per cent such that $\kappa\ll 1$ and $|\gamma|\ll 1$. The shear components can only be estimated from taking the average over a large sample of observed ellipticities $\langle \epsilon_{1,2}\rangle \approx \gamma_{1,2}$. Subsequently, the convergence field can be reconstructed from the shear by performing a KS-inversion as we will explain in subsection \ref{sec:KS}.

In the following subsections, we present the theoretical framework for the summary statistics we use in the tomographic analysis, the probability density function (PDF) of the convergence field, $\kappa$-PDF, and the shear two-point correlation function, $\gamma-$2PCF. 

\subsection{Shear two-point correlation function} 
The real space two-point correlation function ($\gamma-$2PCF) is the most widely used statistic to quantify the shear correlations. It is estimated by multiplying the ellipticities of galaxy pairs and averaging over all pairs separated by an  angular separation $\theta$. As the shear is analogous to a polarisation tensor, it is mathematically convenient to write the shear as a complex quantity, $\gamma = \gamma_1 + i\gamma_2 = |\gamma|\exp(2i\varphi)$ with $\varphi$ the polar angle between the two shear components.

The two components of the $\gamma-$2PCF are defined as 
\begin{equation}
    \xi_{\pm}(\theta) = \langle \gamma \gamma^*\rangle = \langle \gamma_1 \gamma_1\rangle(\theta) \pm \langle \gamma_2 \gamma_2\rangle (\theta) \,.  
    \label{eq:xi}
\end{equation}
As the shear components and the convergence are proportional to second derivatives of the 2D gravitational potential, the two components of the $\gamma-$2PCF are uniquely determined by the angular convergence power spectrum in the following way
\begin{equation}
    \xi_{\pm}(\theta) = \int \frac{\ell d\ell}{2\pi}J_{0,4}(\ell\theta)C_{\kappa\kappa}(\ell) \label{eq:xi}
\end{equation}
where $J_i$ are the spherical Bessel functions, $i=0,4$ for $\xi_{+}$ and $\xi_{-}$ respectively. The convergence power spectrum is given by
\begin{equation} 
    C_{\kappa\kappa}(\ell) = \int^{\infty}_{0} d \chi\, w(\chi(z))^2 P_{\rm NL}(k=\ell/\chi, z)\,,
\end{equation}
and $P_{\rm NL}(k,z)$ is the non-linear matter power spectrum.

\subsection{Convergence one-point PDF} \label{sec:kappa_PDF_th}

This section details the theoretical framework for calculating the weak lensing convergence probability distribution function (PDF) \citep{Barthelemy:2019ciu, Boyle:2020bqn}. We consider a set of source galaxies with a given redshift distribution.

\subsubsection{Predicting $\kappa-$PDF from the matter CGF}

In practice, the probability density function (PDF) is predicted via the cumulant generating function (CGF) as detailed in Appendix \ref{app:CGF}. Both the CGF and PDF are determined within the framework of large deviation theory (LDT) which predicts the distributions of matter density in the limit of small variances as opposed to small density contrasts. The convergence PDF $P(\kappa)$ and the CGF $\phi(\lambda)$ are related through an inverse Laplace transformation in the following way 
\begin{equation}
    P(\kappa) = \int^{+i\infty}_{-i\infty} \, \frac{d\lambda}{2 \pi i} \,  e^{\phi(\lambda)-\lambda \kappa}\,.
\end{equation}
The CGF is the logarithm of the moment generating function and is given by
\begin{equation}
    \phi(\lambda) = \sum_{l=1}^{\infty}\, \langle  \kappa^l \rangle_c \frac{\lambda^l}{l!}\,,
\end{equation}
where the convergence cumulants are defined as 
\begin{equation}
    \langle \kappa^l \rangle_c = \int_{0}^{\chi_s} \, \left(\prod_{i=1}^{l}  d\chi_i\,w(\chi_i) \right) \langle \delta_1\cdots \delta_l\rangle_c\,.
\end{equation}
The cumulants are computed in the small angle approximation which is valid when the transverse distances $\chi|\bm{\theta}_i-\bm{\theta}|$ are much smaller than the radial distance $\chi$. 
Since the correlation length is much smaller than the Hubble scale $c/H(z)$ the variable $\chi_i$ can be written as $\chi_i = \chi_1+r_i $  with $|r_i|\ll c/H(z)$ such that\footnote{This is the Limber approximation which allows to relate the angular and spatial correlation functions by considering averages over cylinders of the same size.}
\begin{align}
    \langle \kappa^l \rangle_c &= \int_{0}^{\chi_s} \, d\chi_1\, w(\chi_1)^l \\
   \notag&\times \int_{-\infty}^{\infty} \left( \prod_{i=2}^{l}\, dr_i \right) \langle \delta(\chi_1\bm{\theta}_1,\chi_1) \cdots \delta(\chi_l\bm{\theta}_l,\chi_1+r_l)\rangle_c\,.
\end{align}
Therefore, assuming $\delta(\chi_l\bm{\theta}_l,\chi_1+r_l)\simeq\delta(\chi_l\bm{\theta}_l,\chi_1)$ and considering filtering effects, the cumulants of the smoothed convergence field are related to $\delta_{\cyl}$, the 3D density contrast average over a cylinder of transverse size $R_{\theta}=\chi\theta$ and length $L$, in the following way
\begin{equation}
\label{eq:cumulants}
\langle \kappa_{\theta}^l \rangle_c = \int_{0}^{\chi_s} \, d\chi\,w(\chi)^l\, \langle \delta_{\cyl}^{l}\left(R_{\theta},L\right)\rangle_c \,L^{l-1} \,.
\end{equation}
The density field appearing in this equation is smoothed with a top-hat filter in cylindrical coordinates as \citep{Uhlemann:2017kvh}
\begin{equation}
    \delta_{\cyl}(\bm x)\!=\!\!\int\!\! d^3\bm x'\, W_{\rm2D}(|\bm{\theta}'-\bm{\theta}|,R_{\theta})W_{\rm 1D}\left(|\chi'-\chi|,\tfrac{L}{2}\right)\delta(\bm{x}')\,,
\end{equation}
with $\bm{x} = \chi(\bm{\theta},1)$, for an infinite long cylinder $L\rightarrow \infty$ the filter reduces to $W_{2D}$.
Hence, the convergence CGF is related to the CGF of the density contrast in cylinders as described in \cite{Barthelemy:2019ciu,Boyle:2020bqn} 
\begin{equation}
    \phi_{\theta}(\lambda) = \int^{\chi_s}_0 \,d\chi\, \lim_{L\rightarrow \infty}\phi_{\theta,\text{cyl}}\left(L w(\chi)\lambda\right)L^{-1}\,.
\end{equation}
We briefly review the procedure to obtain the CGF of density contrast in cylinders using LDT in appendix \ref{app:CGF}. 

\subsubsection{Including shape noise}\label{sec:shape_noise}
Observed galaxy shapes results mainly from two contributions: cosmic shear due to gravitational weak lensing, and intrinsic galaxy ellipticity. The latter, a primary source of noise (shape noise), follows a Gaussian distribution with variance $\sigma_\epsilon^2$. The effect on the PDF due to shape noise is included by performing a convolution with a Gaussian distribution centred at zero and with a variance $\sigma_{\rm SN}^2$ as in \cite{DES:2016wzy,Boyle:2020bqn} 
\begin{equation}
    P_{\noisy}(\kappa) = \frac{1}{\sqrt{2 \pi } \sigma_{\rm SN}} \int d \kappa' \exp\left(-\frac{(\kappa-\kappa')^2}{2\sigma_{\rm SN}^2}\right) P(\kappa')\,.
\end{equation}
 $\sigma_{\rm SN}^2$ is the shape noise variance at a given smoothing scale $\theta$ weighted by the source galaxy density as
\begin{equation}
\label{eq:sigma_noise_theta}
    \sigma_{\rm SN}^2 = \frac{\sigma_{\epsilon}^2}{n_g \Omega_{\theta}} \,,
\end{equation}
with $\Omega_{\theta}$ the solid angle (in arcmin$^2$).

For equi-populated redshift bins, the shape noise contribution is the same across all redshift bins. Hence, the effect on the convergence PDF depends on the lensing projection kernel (equation \ref{eq:weight_function}) for each source redshift bin, i.e on how much the light from the source galaxies is affected by weak lensing. As the projection kernel and its amplitude increase with the redshift of the sources, higher source redshift bins exhibit a stronger lensing signal. Consequently, the impact of shape noise on the PDF is strongest for sources at the lowest redshifts. For a Gaussian shape noise, decorrelated from the signal, it affects only the variance of the PDF such that the variance of the noisy $\kappa-$ PDF is equal to the variance of noise-free $\kappa-$ PDF plus the shape noise variance in the aperture of size $\theta$
\begin{equation}
    \label{eq:PDF_SN_variance}
    \sigma_{\noisy}^{2} = \sigma_{\noisefree}^2 + \sigma_{\rm SN}^2 \,,
\end{equation}
such that the inclusion of shape noise leaves all the higher-order cumulants unchanged and thus produces a more Gaussian convergence field.

Additional systematic effects, like photometric redshift errors, shape calibrations or galaxy intrinsic alignments (IA) caused by external gravitational tidal fields, can be incorporated into the modelling of the $\kappa-$PDF. See Appendix~\ref{app:IA} where we include and validate the impact of IA as described by non-linear alignment models \citep[NLA,][]{Hirata:2007np,Bridle:2007ft} into the $\kappa-$PDF predictions.

\subsubsection{Convergence reconstruction: Kaiser-Squires}\label{sec:KS}
As the convergence field is not a direct observable, it needs to be reconstructed from the shear estimations. To obtain the $\kappa$-maps from the shear ones, we perform a Kaiser-Squires inversion (KS) \citep{Kaiser:1992ps}. To explain the mathematical framework behind the KS inversion, it is convenient to  decompose the convergence field into its electric-like  and magnetic-like  modes as $\kappa = \kappa^E + i \kappa^B$. The shear and the convergence fields are related through a linear combination of equations \eqref{eq:shear_equations} in Fourier space as
\begin{align}
    \hat{\gamma} & = \hat{P}\hat{\kappa}\,,\qquad \qquad \hat{\kappa} = \hat{P}^*\hat{\gamma}\,,
\end{align}
where $\hat{P}$ is a linear operator in Fourier space given by
\begin{align}
   \hat{P} = \frac{k_1^2-k_2^2+2ik_1k_2}{k_1^2+k_2^2} 
\end{align}
with ${k_1,k_2}$ the frequencies associated with the coordinates $\bm{\theta}$ on the orthogonal plane to the light-of-sight.
After applying the operator $\hat{P}$ on the shear field, we get the convergence field by inverting back to real space. As the convergence field is a scalar quantity, we expect the $B$-mode to vanish. However, for finite and masked fields the KS inversion produces spurious frequencies which generates a leakage of information from the $E$-mode to the $B$-mode and produces border effects on the convergence maps. Also, of relevance direct to us, this effect reduces the variance in the convergence field.

Several techniques have been developed to mitigate the impact of reconstruction, such as the inpainting method \citep{Elad:2005cq,Pires:2008dw}. This technique fills the masked regions by extrapolating the information from the observed data. It iteratively finds a sparse representation of data and fills the gaps using this information. Similar methods have been developed,  for example \cite{Remy:2022ixn} uses an iterative Wiener filtering. While these methods minimise the impact of reconstruction, they can not fully recover all lost information, and it is not clear what is the best method since the effectiveness depends on the specific scientific question \citep{DES:2021gua}. 

We can model the loss of information effect into the theory framework. To do this we follow the procedure in \cite{Barthelemy:2023mer} where the theoretical model for the convergence PDF is modified to include the effect of a survey mask. The idea is to consider that we measure the shear values on a  fraction of the sky, where the rest of the sky is masked. For masked fields the shear and convergence correlation functions in Fourier space are affected due to the mixing of frequencies and leakage from E to B modes. This reduction in the power spectrum impacts the shear and convergence variances. In the LDT model, because of the Born and Limber approximations, the convergence is treated as a superposition of independent layers of the matter density field, seen as a succession of lensing events, along the line-of-sight. Therefore, the reconstruction effect can be included at each redshift slice by scaling the non-linear variance of the matter density field according to the modulation. We achieve this by computing the ratio between the true variance and the reconstructed variance of the (fictitiously masked) matter density field. This factor is then used to scale the non-linear variance within the rescaled cumulant generating function (see equation \ref{eq:rescaledCGF}). This method ensures preserved scaling relations between cumulants of the density.

The non-linear variances of the true and  the KS-reconstructed density field are given by \citep{Barthelemy:2023mer, DES:2021lsy} 
\begin{multline}
    \langle \delta_{true}^2 \rangle =  \sum_{\ell} \frac{2l+1}{4\pi \chi(z)^2}W_{\ell}(\chi(z)\theta)^2 P_{\rm NL}(\ell'/\chi(z),z)\,,
\end{multline}
\begin{multline}
    \langle \delta_E^2 \rangle =  \frac{1}{f_{\rm mask}}\sum_{\ell} \frac{2l+1}{4\pi \chi(z)^2}W_{\ell}(\chi(z)\theta)^2f_{\ell}^{-1}\\
    \sum_{\ell'}M_{\ell\ell'}^{EE,EE} P_{\rm NL}(\ell'/\chi(z),z)f_{\ell'}\,,
\end{multline}
where $M_{\ell'\ell}^{EE,EE}$ are the elements of the mode-mixing matrix that contribute to the transfer of information from the unmasked to masked shear $E$-modes. This matrix includes the full complexity of the survey mask, notably the different holes, or disconnected regions. $f_{\ell} = \left[(\ell+2)(\ell-1)\right]/\left[\ell(\ell+1)\right]$ is the normalisation factor in harmonic space for the full sky reconstruction and $W_\ell$ is the smoothing filter in  harmonic space. $f_{\rm mask}$ is a purely geometrical factor defined as the ratio, at a given smoothing scale, of the reconstructed convergence variances respectively taking into account all the pixels on the reconstructed full-sky and only pixels in the field of view. As detailed in \cite{Barthelemy:2023mer}, this factor is a priori independent of cosmology, has a slight smoothing-scale dependence but is directly measurable on actual data. Nevertheless, its value is by construction very close to the sky fraction, $f_{\text{sky}} \simeq A/4\pi$ with $A$ the area of the unmasked region. We thus fix  $f_{\rm mask} = f_{\text{sky}}$ in the remainder of this work. This will be shown to be accurate enough while having no influence on the forecasts we perform.

\subsection{Tomographic strategy}
\begin{figure}[h]
    \centering
    \includegraphics[scale=0.35]{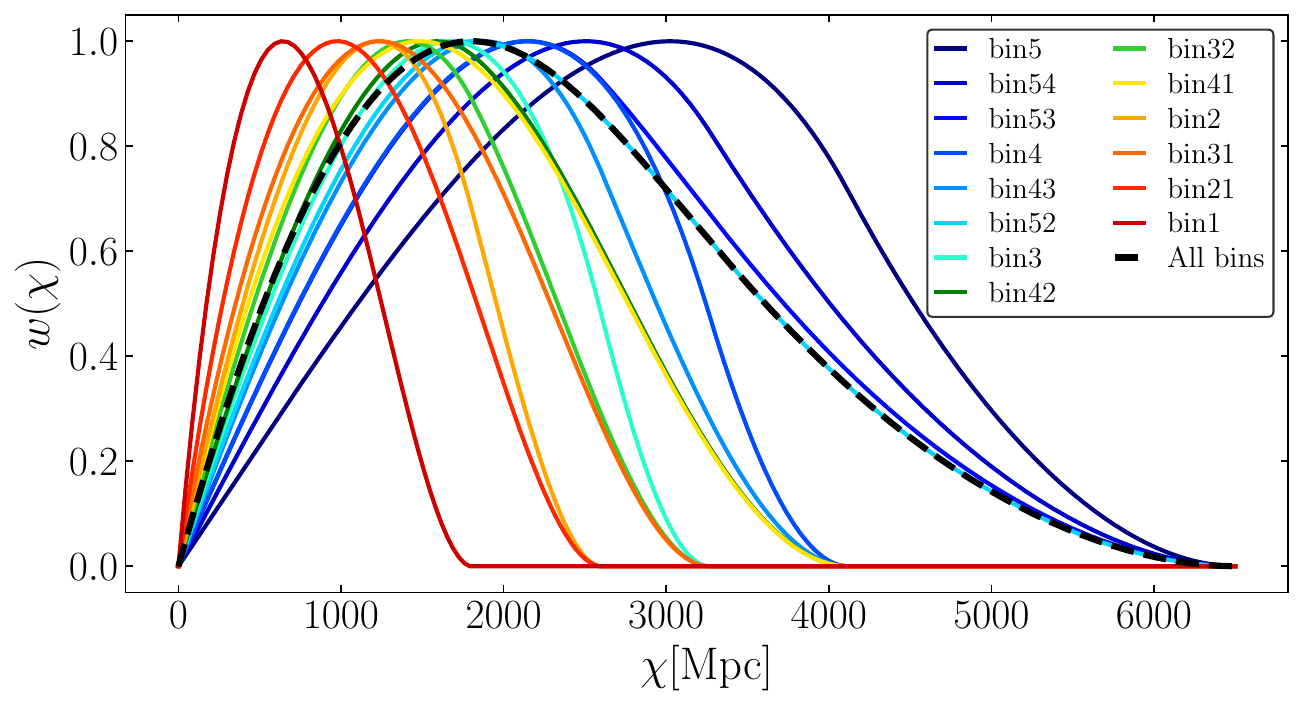}
    \caption{ \label{fig:weak_lens}  Normalised weak lensing projection kernel as a function of the comoving distance for different redshift combinations of the StageIV-like  source redshift distribution.}    
\end{figure}

For  our analysis, we consider a StageIV-like redshift distribution, plotted in figure \ref{fig:n_of_z}. In principle, we divide the redshift distribution into 5 redshift bins. To mitigate the effect on shape noise in the lower redshift bins and to have a fair comparison between the $\kappa$-PDF and the $\gamma$-2PCF our PDF data vector includes the information of 14 redshift combinations. Besides the $\kappa$-PDF for the full redshift distribution and for the individual redshift bins we combine consecutive redshift bins into quartets, triplets and pairs. In table \ref{tab:Redshift-bins} we show the 15 redshift combinations for the $\kappa-$PDF we can form following this strategy, we do not use the lowest redshift bin as it is not accurately predicted by theory in the noise-free case and is highly affected by shape noise. The names on the second column bin$ijkl$ stand for the redshift bins that has been combined for the $\kappa-$PDF. For the $\gamma-$2PCF we consider standard auto and cross-correlations. In figure \ref{fig:weak_lens} we show the normalised weak lensing projection kernel (equation \ref{eq:weight_function}) for the different redshift sources we consider. 
\begin{table}
\begin{centering}
\begin{tabular}{cccccc}
\toprule 
\multirow{2}{*}{} & \multirow{2}{*}{$z_{P(\kappa)}$} & \multirow{2}{*}{$n_{g}\left[\text{arcmin}^{-2}\right]$} & \multicolumn{3}{c}{$\sigma^2$ $[10^{-5}]$}\tabularnewline
\cmidrule{4-6} \cmidrule{5-6} \cmidrule{6-6} 
 &  &  & $\sigma_{\rm SN}^{2}$ & $\sigma_{\text{noisefree}}^{2}$ & $\sigma_{\text{noisy}}^{2}$\tabularnewline
\midrule
\midrule 
$1$ & All bins & 30 & $0.319$ & $2.42$ & $2.74$\tabularnewline
\midrule 
$2$ & bin$2345$ & 24 & $0.398$ & $3.52$ & $3.91$\tabularnewline
$3$ & bin$1234$ & 24 & $0.398$ & $1.66$ & $2.06$\tabularnewline
\midrule
$4$ & bin$345$ & 18 & $0.531$ & $4.71$ & $5.24$\tabularnewline
$5$ & bin$234$ & 18 & $0.531$ & $2.57$ & $3.10$\tabularnewline
$6$ & bin$123$ & 18 & $0.531$ & $1.14$ & $1.68$\tabularnewline
\midrule
$7$ & bin$45$ & 12 & $0.797$ & $6.28$ & $7.08$\tabularnewline
$8$ & bin$34$ & 12 & $0.797$ & $3.46$ & $4.25$\tabularnewline
$9$ & bin$23$ & 12 & $0.797$ & $1.96$ & $2.76$\tabularnewline
$10$ & bin$12$ & 12 & $0.797$ & $0.70$ & $1.50$\tabularnewline
\midrule
$11$ & bin5 & 6 & $1.59$ & $8.98$ & $10.60$\tabularnewline
$12$ & bin4 & 6 & $1.59$ & $4.47$ & $6.07$\tabularnewline
$13$ & bin$3$ & 6 & $1.59$ & $2.70$ & $4.30$\tabularnewline
$14$ & bin2 & 6 & $1.59$ & $1.45$ & $3.05$\tabularnewline
$15$ & bin$1$ & 6 & $1.59$ & $0.30$ & $1.89$\tabularnewline
\bottomrule
\end{tabular}
\par\end{centering}
\caption{\label{tab:Redshift-bins} Slicing of the StageIV-like redshift distributions. Number density of galaxies on each redshift bin combinations. Shape noise variance (from equation \ref{eq:sigma_noise_theta}), variance of the noise-free PDF (measured directly on the noise-free PDF) and variance of the Noisy PDF (from equation \ref{eq:PDF_SN_variance}) for a smoothing scale $\theta=15$ arcmin.} 
\end{table}

\section{Theory validation with SLICS simulations} \label{sec:theory_validation}
In this section, we validate the theoretical frameworks for the summary statistics, the $\kappa-$PDF and $\gamma-$2PCF, with measurements from numerical weak lensing  simulations. We describe these here, and detail our methods to measure the shear and convergence statistics from the corresponding shear and convergence maps. Finally, we compare the theoretical predictions against measurements for our summary statistics.   

\subsection{Weak lensing simulations}
We use the SLICS (Scinet LIghtcone Cosmological Simulations)  \citep{Harnois-Deraps:2018bcv} and {\it cosmo}-SLICS \citep{Harnois-Deraps:2019rsd} weak lensing simulations to estimate the covariance matrix and validate our theoretical framework. Both are dark matter-only simulations obtained by using the gravity solver {\sc cubep$^3$m} \citep{Harnois-Deraps:2012qtf}. The comoving side length of the simulation boxes is $L_{\text{box}}=505h^{-1}\text{Mpc}$, from which  light-cones are extracted with an area of $A_{\rm SLICS} = 100 \text{deg}^2$. The SLICS simulation box contains $1536^3$ particles each with mass $m_p=2.88\times10^9 h^{-1}\text{M}_{\odot}$ resolving dark matter halos with masses below $10^{11}h^{-1}\text{M}_{\odot}$ and accurately simulates structure formation well into the non-linear regime. The SLICS cosmology is determined by the best-fitting WMAP9+BAO+SN parameters \citep{WMAP:2012nax} as presented in Table \ref{tab:SLICS}. The {\it cosmo}-SLICS suite is available for 25 distinct cosmologies, organised in a Latin hypercube configuration, varying $\Omega_{\rm m}$, $h$, $S_8=\sigma_8 \sqrt{(\Omega_{\rm m}/0.3)}$ and $w_0$ and with the particle mass varying with $\Omega_{\rm m}$ and $h$ in the range $[1.42, 7.63] \times 10^9\text{M}_{\odot}$. For each of these cosmologies, there are a total of $10\times2$ realisations, 10 pairs with initial conditions carefully selected to ensure that the initial power spectrum of each pair averages to the theoretical prediction \citep{Harnois-Deraps:2019rsd}, effectively minimising the impact of cosmic variance \citep{Harnois-Deraps:2019rsd}.

A shape noise field $\boldsymbol{\epsilon}$ is infused on the shear catalogues by sampling a Gaussian distributions with standard deviation $\sigma_\epsilon$ (per component), constructing observed  ellipticities $\boldsymbol{\epsilon}_{\rm obs}$  as:
\begin{align}
{\boldsymbol \epsilon}_{\rm obs} = \frac{\boldsymbol{ \gamma} + \boldsymbol{ \epsilon}}{1 + \boldsymbol{\gamma}^* \boldsymbol{\epsilon}} \, ,
\end{align}
where bold symbols refer to complex notation, i.e. $\boldsymbol{\gamma} = \gamma_1$ + i$\gamma_2$.
We use the  SLICS and  {\it cosmo}-SLICS \textit{StageIV-like} simulation suites for which the source galaxy number density ($n_g=30 \text{arcmin}^{-2}$), galaxy shape noise ($\sigma_{\epsilon}=0.26$), and redshift distribution (FIG \ref{fig:n_of_z}) are representative of upcoming lensing surveys.
\begin{table}
\begin{centering}
\begin{tabular}{ccccccc}
\toprule 
$\Omega_{\rm m}$ & $\Omega_{\Lambda}$ & $\Omega_{b}$ & $h$ & $\sigma_{8}$ & $n_{s}$ & $w_0$\tabularnewline
\midrule
\midrule 
$0.2905$ & $0.7095$ & $0.0473$ & $0.6898$ & $0.826$ & $0.969$ & -1.0\tabularnewline
\bottomrule
\end{tabular}
\par\end{centering}
\caption{\label{tab:SLICS} Cosmological parameters of the SLICS simulations}.
\end{table}

\subsection{Shear and convergence maps }
We use the SLICS shear catalogues available for $N_{\rm SLICS}=923$ realisations and the {\it cosmo}-SLICS catalogues for the 20 realisations to generate  noise-free and noisy shear maps. Each shear catalogue contains information about the positions $(x,y)$, the redshift $z$, shear values $(\gamma_1,\gamma_2)$, and observed ellipticities $(e_{obs,1},e_{obs,2})$. We obtain the shear and observed ellipticity maps by projecting the shear or ellipticity values for each galaxy onto a  $2D$ Cartesian grid. To ensure we have at least one galaxy per pixel, each map has a size $600^2$ pixels. The value on each pixel is the average shear or ellipticity assigned to the corresponding pixel\footnote{We use the Pylians libraries \citep{Pylians} \href{https://pylians3.readthedocs.io/en/master/index.html}{https://pylians3.readthedocs.io/en/master/index.html}}.  

\begin{figure}
\centering
    \includegraphics[scale=0.6]{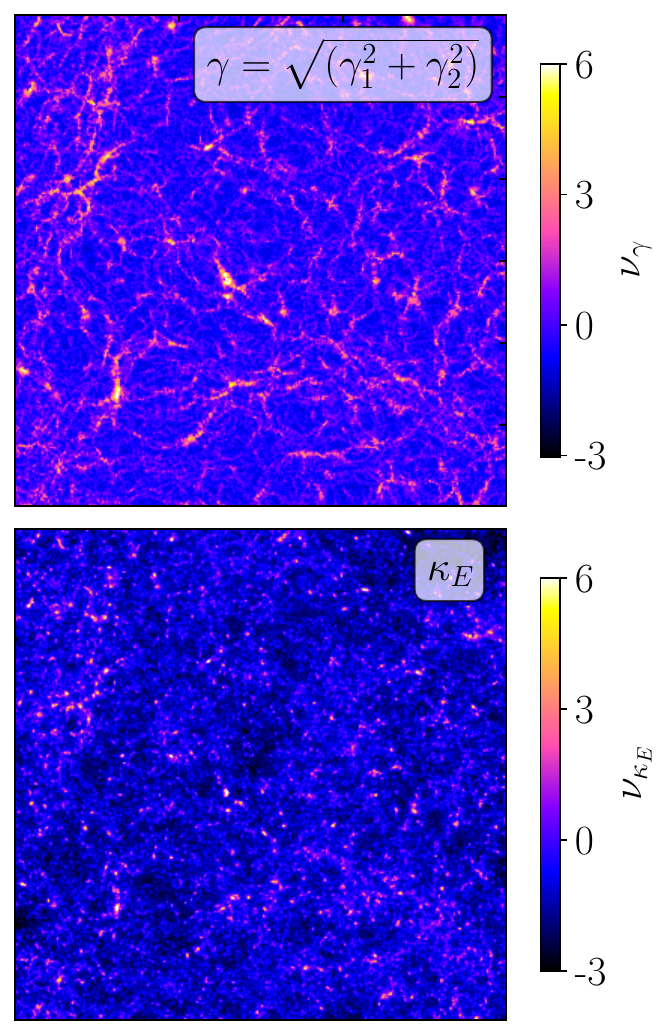}
    \caption{Weak lensing maps. Upper: Absolute value of the shear field. Lower: KS-reconstructed convergence map. The background colour represents the fluctuation field in units of the field  standard deviation, e.g $\nu_{\gamma} = (\gamma - \langle \gamma \rangle)/ \sqrt{\langle \gamma^2\rangle_c}$.}
    \label{fig:maps_no_noise}
\end{figure}

From the shear maps we reconstruct the convergence maps by using the KS-inversion presented in section \ref{sec:KS}\footnote{We use bin2d, ks93 functions from lenspack  \href{https://github.com/CosmoStat/lenspack/tree/master/lenspack}{https://github.com/CosmoStat/lenspack/tree/master/lenspack}}.  In Figure \ref{fig:maps_no_noise} we present the weak lensing maps for the non-tomographic case. The upper panel shows the map obtained by combining the two components of the shear as,  $\gamma = \sqrt{\gamma_1^2+\gamma_2^2}$ and the lower panel is the KS-reconstructed kappa map $\kappa_E$. In both cases the colour represents the size of the fluctuation field in units of the field  standard deviation, e.g. $\nu_{\gamma} = (\gamma - \langle \gamma\rangle)/\sqrt{\langle \gamma^2\rangle_c}$.

\subsection{$\kappa-$PDF validation}\label{sec:validation}

We measure the $\kappa-$PDF on the 923 KS-reconstructed SLICS convergence maps for three smoothing scales,  $\theta=\{10,15,20\} \text{ arcmin}$ obtained by performing a convolution with a top-hap filter with a radius equal to $\theta$. We measure the PDF by computing a histogram of the smoothed $\kappa-$maps, for each redshift combination, and we average over the realizations to obtain the measured  $\kappa-$PDF. 

Theoretical predictions presented in section \ref{sec:kappa_PDF_th} have been implemented in the  public code \cosmomentum\footnote{\href{https://github.com/OliverFHD/CosMomentum}{https://github.com/OliverFHD/CosMomentum}} \citep{Friedrich:2019byw}. We use \cosmomentum to obtain the theoretical predictions for the SLICS and {\it cosmo}-SLICS cosmologies and the redshift distribution for each of the 15 redshift combinations. We include the effect of shape noise by convolving the PDF with a Gaussian random field with $\sigma^2_{\rm SN}$ variance corresponding to each redshift bin, as explained in section \ref{sec:shape_noise}. The non-linear power spectrum is modelled using the fitting function \halofit \cite{Takahashi:2012em}. The variance for the KS-reconstructed convergence field is obtained by using the public code \ldt\footnote{ \href{https://github.com/AlexandreBarthelemy/L2DT-Lensing-with-Large-Deviation-Theory}{https://github.com/AlexandreBarthelemy/L2DT}} \citep{Barthelemy:2023mer}. 

\begin{figure}[h]
    \centering
    \includegraphics[width=0.45\textwidth]{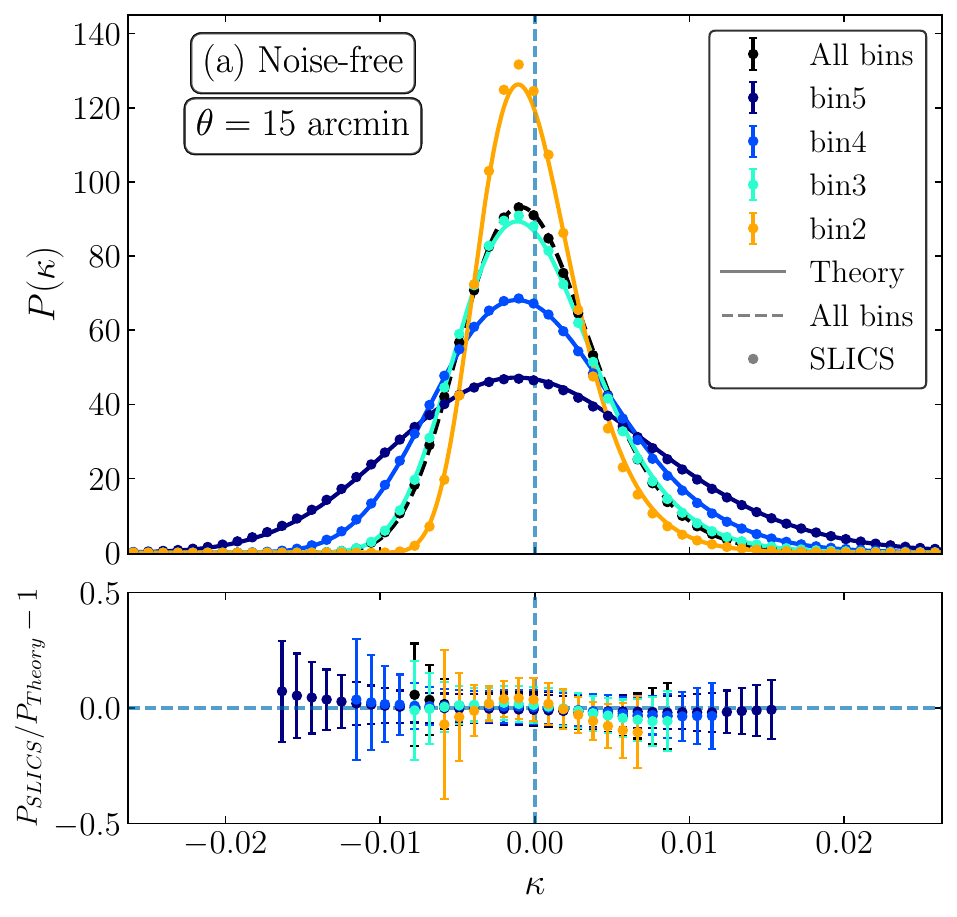}
    \includegraphics[width=0.45\textwidth]{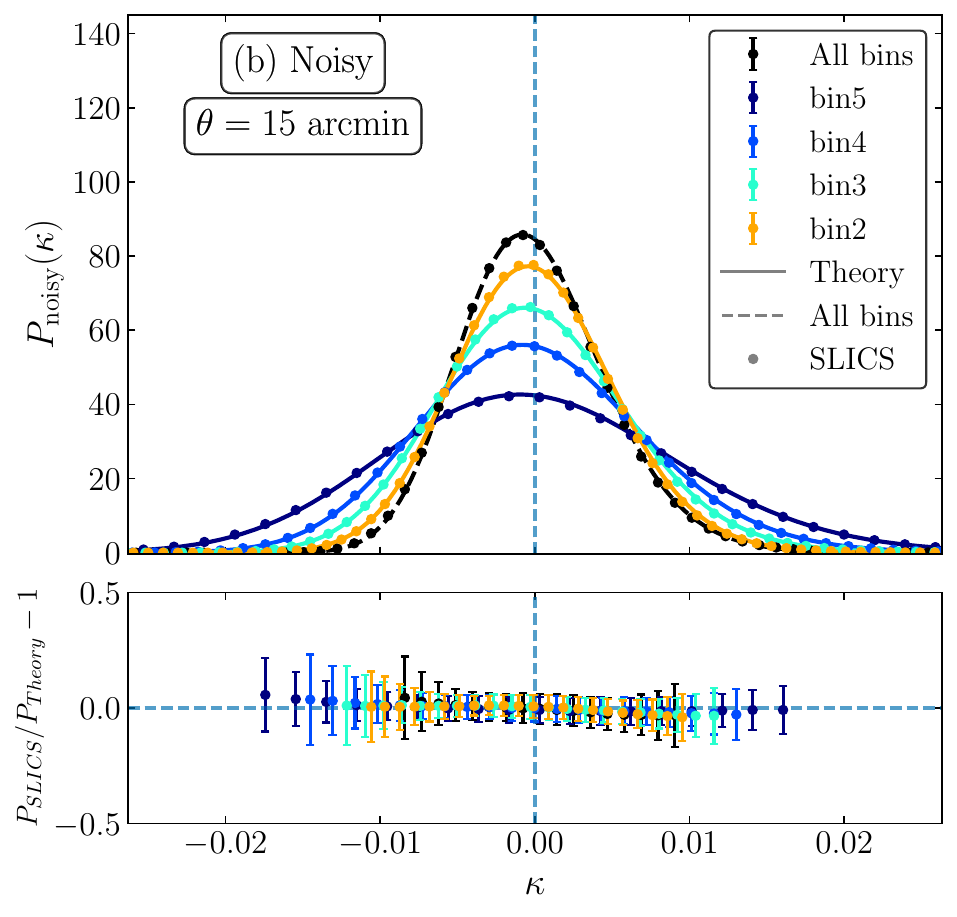}
    \caption{$\kappa-$PDF theory validation with SLICS simulations for $\theta=15$ arcmin. (a) Noise-free. (b) Noisy. Upper panels: $\kappa$-PDF comparison between the theory (solid and dashed lines) and the SLICS measurements (with the error bars given by standard deviation of the mean) for 4 auto  redshift bins and all bins combined. Lower panels: Residuals between the measure SLICS and the predicted $\kappa$-PDF within the $2\sigma$ information around the peak, the error bars represent the standard deviation over realizations. Other tomographic bin combinations show a similar agreement (see figure \ref{fig:noisy_kappa_PDF_SLICS}).}
    \label{fig:freenoisy_vs_noisy_PDF_validation}
\end{figure}

In figure \ref{fig:freenoisy_vs_noisy_PDF_validation}, we compare $\kappa-$PDF measurements with theoretical predictions for $5$ selected redshift combinations for (a) the noise-free convergence, and (b) the convergence in the presence of shape noise. On the upper panels the SLICS measurements are presented with data points, we observe that the theoretical $\kappa-$PDF (solid and dashed lines) predict the measurements with high accuracy. On the lower panels, we plot the residuals between the measured SLICS and the predicted $\kappa-$PDF. The error bars represent the standard deviation across realisations, which is of the order of $10\%$\footnote{The error on the mean is a factor of $\sqrt{N_{\rm sim}-1}\approx 30$ smaller.}. When rescaled to the full survey area, the error bars shrink to around $1\%$ precision. For the noise-free $\kappa$, the PDF is less accurate for the smaller redshift bins, partially due to an overprediction of the variance that leads to a lower peak height. This is a direct consequence of the fact that we can't directly measure $f_{\rm mask}$ with only simulated small patches, and that we thus fixed it to the $f_{\rm sky}$ value. This effect is mixed with inherent inaccuracies of the non-linear matter power spectrum recipe that we use, i.e. \halofit. Additionally, smaller redshift bins probe smaller non-linear scales due to the mixing of scales inherent to weak lensing for which the mildly nonlinear predictions are not ideally suited. Nevertheless, predictions within the 2-$\sigma$ region around the PDF peak are accurate at the 5\%-level for bin2 and below that for higher redshift bins. We find that, including the effect of the KS-reconstruction by rescaling the non-linear  variance at the CGF level is enough to obtain good agreement between the theoretical predictions and measurements, in both the noise-free and noisy cases. 

\begin{figure*}[ht]
    \centering
    \includegraphics[width=0.8\textwidth]{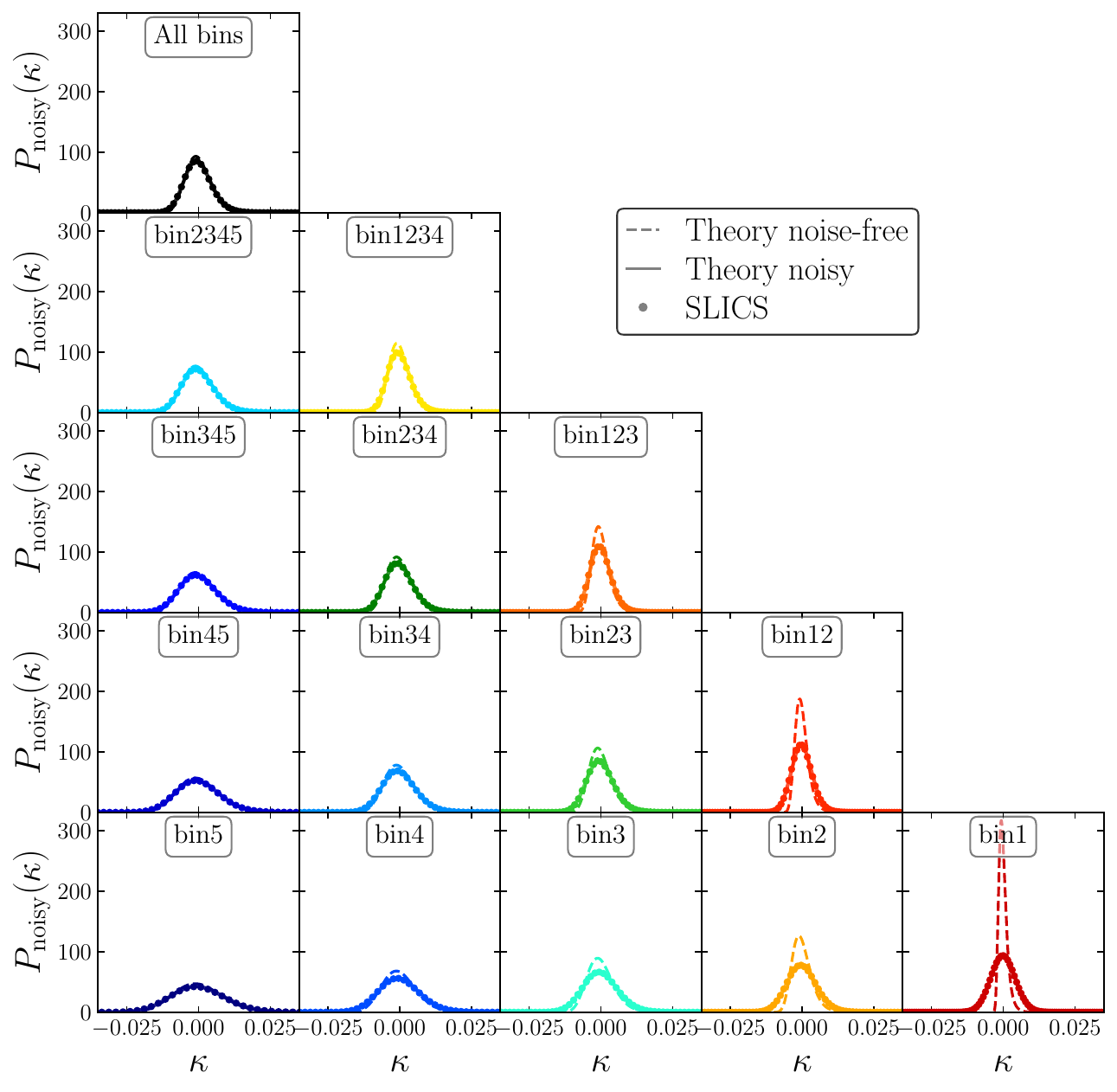}
    \includegraphics[width=0.8\textwidth]{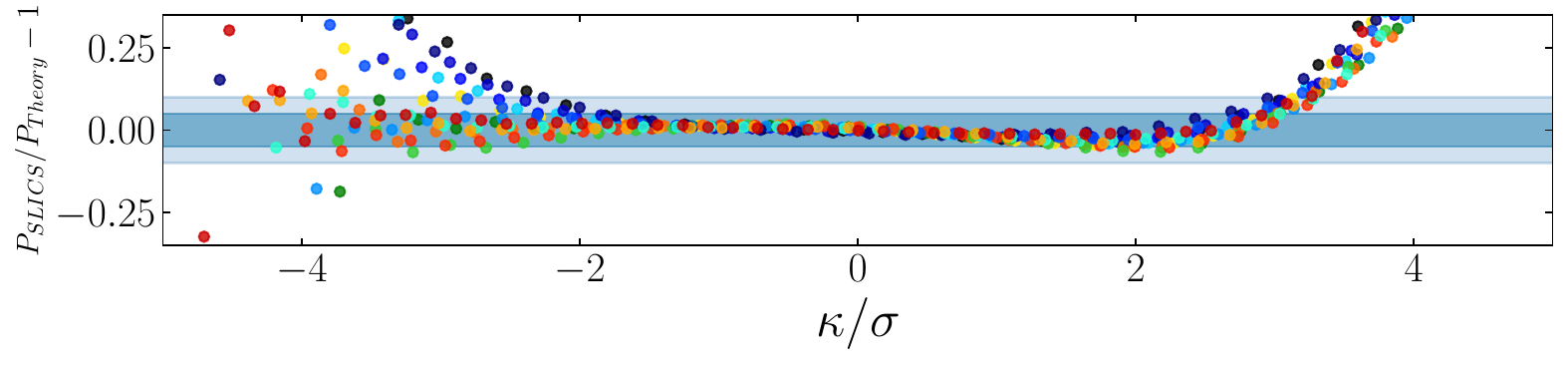}
    \caption{Noisy $\kappa$ PDF validation with SLICS simulations for the 15 redshift combinations in our tomographic strategy. \textbf{Upper panel:} comparison between the theory predictions (solid line) and  SLICS measurements (dots with error bar given by the standard deviation of the mean). Theory predictions for the noisy-free $\kappa-$PDF (dashed lines) are also shown. \textbf{Lower panel:} Residuals between the SLICS measuremets and the theory predictions, the shaded bands represent the $5\%$ (dark) and the $10\%$ (light) errors. Same colour scheme from upper panel.}
    \label{fig:noisy_kappa_PDF_SLICS}
\end{figure*}

In Figure \ref{fig:noisy_kappa_PDF_SLICS} we validate the full set of theoretical predictions (solid lines) for the noisy $\kappa$-PDF by comparing them with  measurements (dots) from the noisy SLICS KS-reconstructed convergence field, for the 15 redshift combination from our tomographic set. We find good agreement between theory and measurements for all redshift combinations. In order to see the effect of shape noise on different redshift combinations, we also include theoretical predictions for the noise-free $\kappa-$PDF (dashed lines). We can see clearly that lower redshift bins are more affected by shape noise. On the lower panel, we present the residuals between the SLICS measurements and the noisy theoretical predictions, here we have normalised the $\kappa-$values by the width $\sigma$ of the $\kappa-$PDF. Again, the percentage errors are within the $5\%$ around the peak, within the $2\sigma$ of the PDF, for all redshift combinations. We can see from the residuals outside the $2\sigma$ region around the peak that the theoretical predictions for the  tails of the PDF are less accurate. In consequence, for our Fisher analysis (next section) we cut the tails by keeping the information within the $2\sigma$ of the PDF, that still has enough non-Gaussian information while admitting a Gaussian likelihood analysis. Probing further into the tails would require refining our theoretical methods with nulling techniques to limit the range of physical scales probed \citep{Barthelemy:2019ciu}, or accounting for beyond leading-order gravitational effects from spherical collapse \citep{Valageas:2001tails} and aspherical corrections \citep[so far only computed for the matter PDF in][]{Ivanov2019}. Another avenue could be more phenomenological and could consist in including effects from the 1-halo regime \citep{Thiele:2020rig,Kainulainen_2011}.

\subsection {$\gamma-$2PCF validation}

We measure the $\gamma-$2PCF  $\xi_{\pm}(\theta)$ in 10 logarithmically spaced bins from separations $\theta$ from 1 to 600 arcmin using \treecorr\footnote{\href{https://github.com/rmjarvis/TreeCorr}{https://github.com/rmjarvis/TreeCorr}} \citep{Jarvis_2004_treecorr,Jarvis2015_treecorr}.   The tomographic data vector is built from the usual auto-correlation of individual redshift bins and cross-correlations between two redshift bins. Theoretical predictions for the $\gamma-$2PCF are obtained with \pyccl\footnote{\href{https://github.com/LSSTDESC/CCL}{https://github.com/LSSTDESC/CCL}} \citep{LSSTDarkEnergyScience:2018yem} using the described source redshift distribution $n(z)$ and corresponding cosmologies. We use the CAMB setting with the `original' \halofit \citep{Smith_2003_halofit} to ensure compatibility with \cosmomentum currently relying on that incarnation.  We have checked consistency between the predictions obtained from \pyccl and CosmoSIS \citep{Zuntz:2014csq}.

\begin{figure*}[ht]
    \centering
    \includegraphics[width=0.8\textwidth]{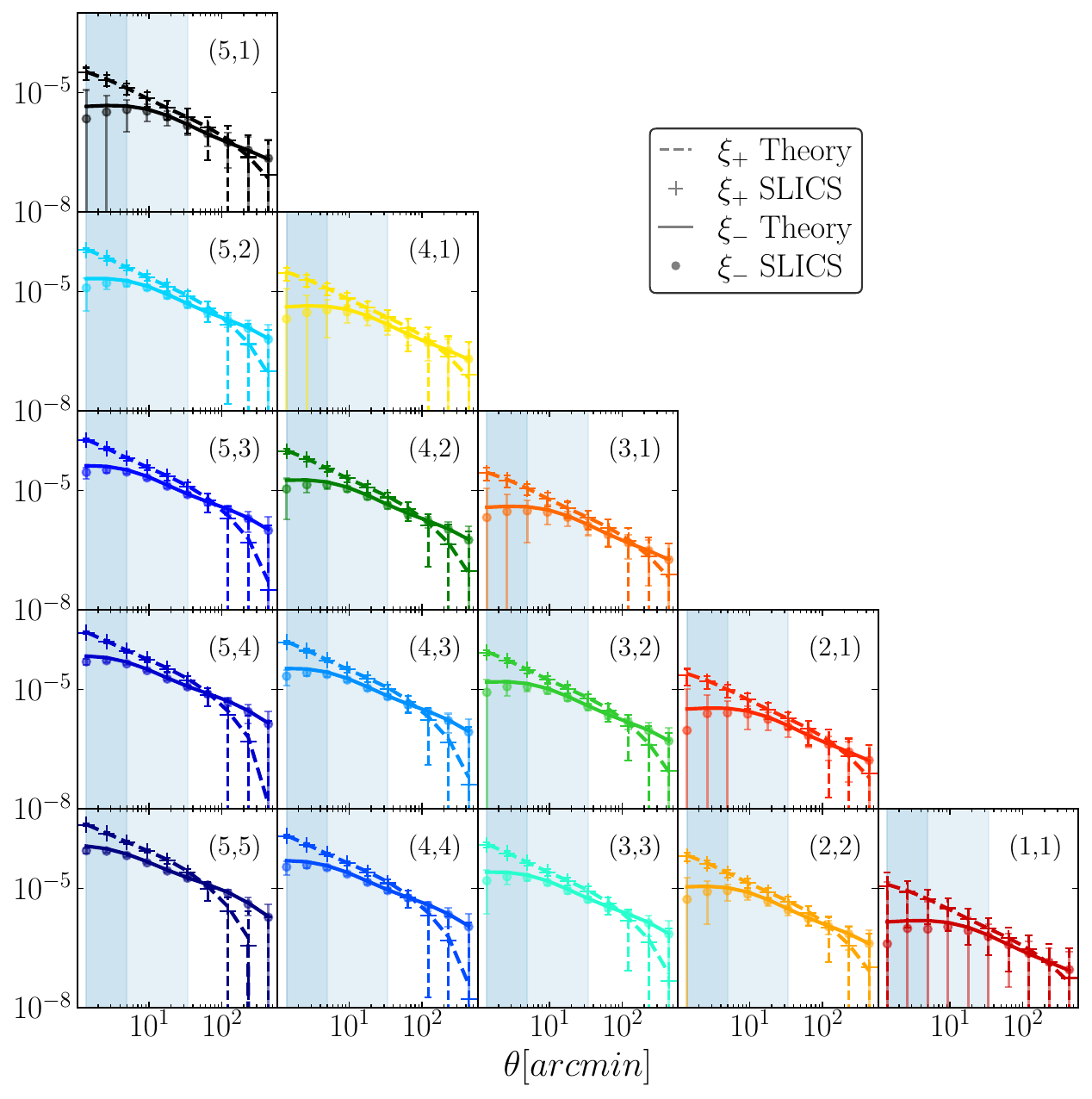}
    \caption{$\gamma-$2PCF  validation with SLICS simulations for the individual redshift bins and cross-correlations between two redshift bins. Comparison between theory predictions  for the two components of the $\gamma-$2PCF (dashed and solid lines for $\xi_{+}$ and $\xi_{-}$ respectively) and the $\gamma-$2PCF SLICS measurements from the noisy shear maps (crosses and dots for $\xi_{+}$ and $\xi_{-}$ respectively). The error bars are obtained from the standard deviation over realisations (light dashed and solid dark lines  for $\xi_{+}$ and $\xi_{-}$ respectively). The shaded regions represent the scales cuts we will use for the Fisher joint analysis (dark and light for $\xi_{+}$ and $\xi_{-}$ respectively).}
    \label{fig:xi_th_SLICS}
\end{figure*}

Figure \ref{fig:xi_th_SLICS} shows the comparison between the theory and SLICS measurements for the two components of the $\gamma-$2PCF, $\xi_{\pm}$ (see equation \ref{eq:xi}). Predictions for $\xi_{+}$ are accurate for all scales across all redshift combinations, while the predictions for $\xi_{-}$ are less accurate at smaller scales. The difference between the two components $\xi_{\pm}$ is determined by the  Bessel filters in equation \ref{eq:xi}. For small argument $J_0$ is positive while $J_4$ is zero. For the $\gamma-$2PCF the shape noise affects the size of error bars, and more generally its covariance matrix, while the signal remains unchanged. Indeed, shape noise as described in equation \eqref{eq:sigma_noise_theta} has no spatial correlation which equates to a constant noise power spectrum and thus only a contribution to the zero-point contribution to the shear correlation function.

\section{Fisher Forecast} \label{sec:Fisher}
We perform a Fisher forecast to quantify how well the one-point PDF and the combination with two-point correlations can constrain the cosmological parameters $p_i=\{\Omega_{\rm m},S_8 ,w_0\}$ when adding the tomographic information. The Fisher information is given by the Fisher matrix defined as 
\begin{equation}
\label{eq:Fisher_matrix}
    F_{ij} = \frac{\de \bm{D}^{T} }{\de p_i}C^{-1}_{ij}\frac{\de \bm{D}}{\de p_j}\,,
\end{equation}
where $\bm{D}$ is the data vector built from the theoretical predictions for the $\kappa-$PDF and the $\gamma-$2PCF, and $C$ is the data covariance matrix estimated from the measurements of the SLICS simulations. The parameter covariance is obtained from inverting the Fisher matrix.

We will perform various Fisher Forecasts to quantify
\begin{itemize}
    \item the improvement of constraints by adding tomography in the $\kappa-$PDF, see figure~\ref{fig:Fisher_tomo_no_tomo}
    \item the improvement of combining the tomographic $\kappa-$PDF and $\gamma-$2PCF, see figure~\ref{fig:fisher_joint_pdf_2pcf}. 
    \item the effect of varying the number of parameters to forecast, see figures \ref{fig:fisher_joint_pdf_2pcf_with_h} and \ref{fig:fisher_joint_pdf_2pcf_now}.
\end{itemize}

For the $\kappa-$PDF we only use the information in the bulk, keeping the $2\sigma$ around the peak, thus excluding the tails. The 2$\sigma$ region around the peak corresponds to removing the lowest $1.5\%$ and the highest $3.5\%$ $\kappa$ bins, which can be easily identified using the cumulative distribution function as integral of the PDF. We combine the $\kappa-$PDF at three different smoothing scales, $\theta= \{10,15,20\}$ arcmin.  We finally interpolate $N_s = 10$ data points within the corresponding $\kappa$ range. To be consistent with the scales we are probing with the $\kappa-$PDF and taking into account the theoretical accuracy in the $\gamma-$2PCF in the combined Fisher analysis, for $\xi_{+}$ and $\xi_{-}$ we consider scales $\theta>10$ arcmin and $\theta>50$ arcmin, respectively. An individual Fisher forecast for just the $\gamma-$2PCF with  tomography is shown in Appendix \ref{app:xi}.

\subsection{Data vector derivatives}\label{sec:derivatives}
The derivatives of our data vector $\bm{D}$ with respect to the cosmological parameters are obtained by concatenating the derivatives for each summary statistics and each redshift combination that we consider in the analysis. We compute derivatives from the theoretical predictions of the $\kappa-$PDF and the $\gamma-$2PCF using finite differences as
\begin{equation}
    \frac{\de \bm{D}}{\de p_i} =  \frac{\bm{D}(p_i^+)-\bm{D}(p_i^{-})}{p_i^+-p_i^-}\,.
\end{equation}
We take variations of the cosmological parameters $p'=\{\Omega_{\rm m}$,$\sigma_8$,$w_0\} $ around the fiducial SLICS cosmology (see table  \ref{tab:SLICS}). We  increase and decrease the values of $\Omega_{\rm m}$ and $\sigma_8$ by $\sim10\%$ and for $w_0$ by $50\%$ with respect to their fiducial values with equidistant steps. In table \ref{tab:D_parameter} we present the fiducial values, the increment/decrement ($\pm$) and the corresponding values for the individual parameters ($p_i^+,p_i^-$).    
\begin{table}
\begin{centering}
\begin{tabular}{cccc}
\hline 
 & $\Omega_{\rm m}$ & $\sigma_{8}$ & $w_{0}$\tabularnewline
\hline 
\hline 
 \fid & $0.290$ & $0.826$ & -$1.0$\tabularnewline
\hline 
\hline 
$\pm$ & $0.030$ & $0.075$ & $0.50$\tabularnewline
\hline 
$p^{-}$ & $0.260$ & $0.751$ & $-1.50$\tabularnewline
\hline 
$p^{+}$ & $0.320$ & $0.901$ & $-0.50$\tabularnewline
\hline 
\end{tabular}
\par\end{centering}
\caption{\label{tab:D_parameter}Variation of the cosmological parameters.}
\end{table}

To obtain derivatives for the set of parameters $p=\{\Omega_{\rm m},S_8,w_0\}$ from the derivatives of the set $p'=\{\Omega_{\rm m},\sigma_8,w_0\}$ we use the chain rule
\begin{equation}
    \frac{\de \bm{D}}{\de p_i} = \sum_{j} \frac{\de \bm{D}}{\de p_j'}\left(\frac{\de p_j'}{\de p_i}\right)_{\fid}\,,
    \label{eq:chainrule}
\end{equation}
where the subscript $\fid$ means we evaluate the partial derivative at the fiducial values. The transformations on the derivatives~\eqref{eq:chainrule} are equivalent to a transformation of the Fisher matrix as $F_{ij}=M^{T}F'_{ij}M^T$ \citep{Coe:2009xf}, where the matrix transformation is given by \citep{Euclid:2023uha} 
\begin{equation}
 M=\left(\begin{array}{ccc}
1 & 0 & 0\\
 -\frac{\sigma_{8}}{2 \Omega_{m}} & \left(\frac{0.3}{\Omega_{m}}\right)^{1/2} & 0\\
0 & 0 & 1
\end{array}\right)\,.
\end{equation}

\begin{figure}
    \centering
    \includegraphics[width=0.45\textwidth]{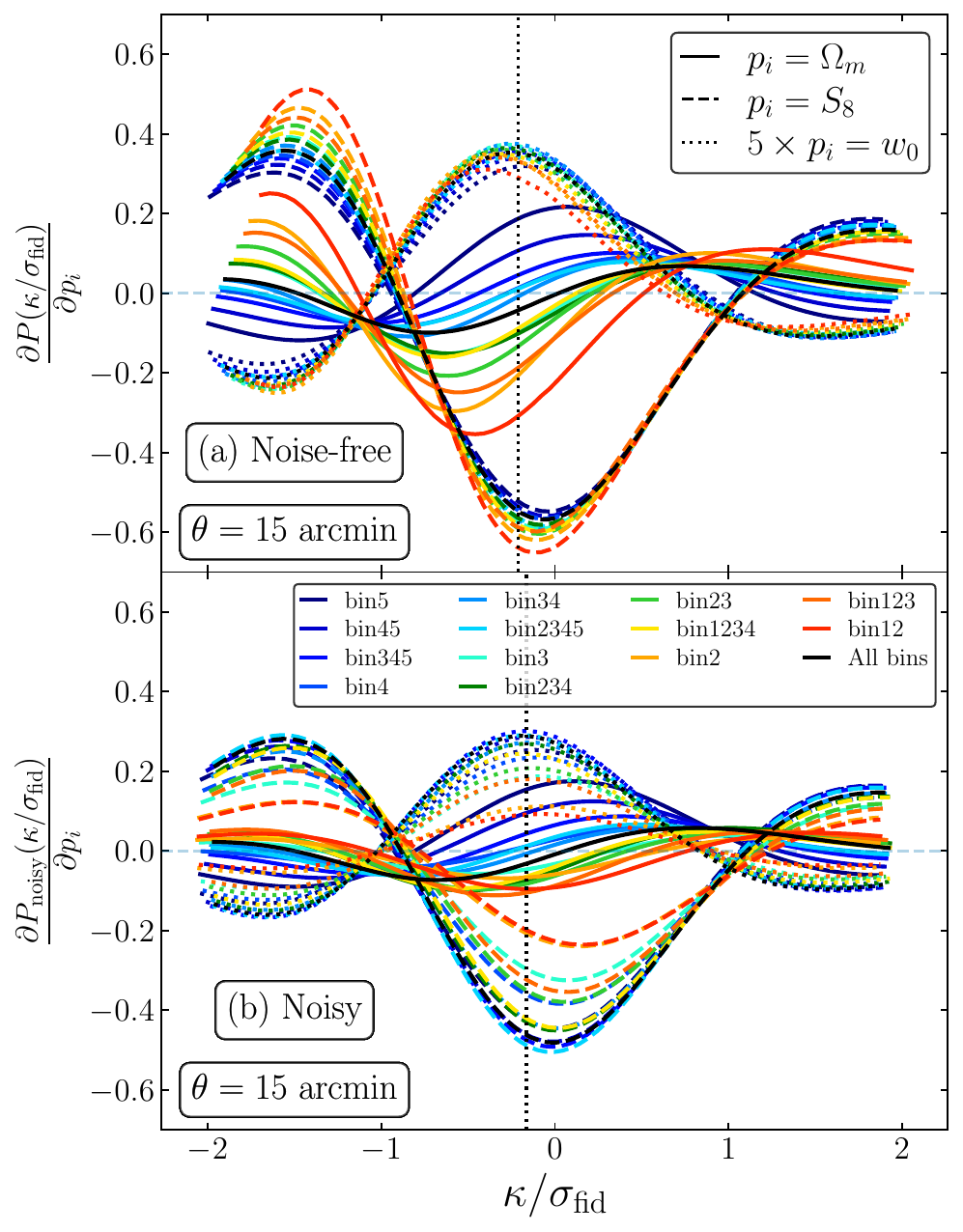}
    \caption{PDF derivatives. The upper and bottom panels show derivatives in the noise-free and noisy cases respectively. We show $\Omega_{\rm m}$ (solid lines), $S_8$ (dashed lines) and $w_0$ (dotted lines) derivatives, where the derivatives for $w_0$ are enhanced by a factor of $5$ for better visualisation. The black dotted vertical line represent the maximum PDF value for the fiducial cosmology for the ``All bins" case.}
   \label{fig:PDF_derivatives} 
\end{figure}

Figure \ref{fig:PDF_derivatives} shows the $\kappa-$PDF derivatives for the $14$ redshift combinations (same colour-code as in figure \ref{fig:weak_lens}, low redshifts in red to high redshifts in blue) and a smoothing scale of $\theta= 15$ arcmin. We show the derivatives of the $\kappa-$PDF normalised by the width $\sigma_{\rm fid}$ of the fiducial $\kappa-$PDF, such that we can visualise the $2\sigma$ information around the bulk for the different redshift bins. $w_0$ derivatives are scaled by a factor of $5$ for a better visualisation. The upper (a) and bottom (b) panels shows derivatives for the noise-free and noisy $\kappa-$PDF respectively, where the dotted vertical line represent the most probable $\kappa-$value for the fiducial cosmology for the non-tomography redshift distribution (“All bins”). 

We observe that derivatives with respect to $S_8$ and $w_0$ are almost symmetric around the dotted vertical line. This behaviour is caused because variations with respect to $\sigma_8$ and $w_0$  mainly impact the variance of the PDF. Note  that the signatures of $S_8$ and $w_0$ in the $\kappa-$PDF are significantly less degenerate than they are for the $\gamma-$2PCF derivatives shown in figure~\ref{fig:xi_derivatives}. Increasing (decreasing) the value of $\sigma_8$ increase (decrease) the non-linear variance, which translates in a PDF with larger (smaller) variance and then a lower (higher) peak. Additionally,  increasing (decreasing) the variance through  $\sigma_8$ also shift to the left (right) with respect to the fiducial PDF, as the skewness increases with the variance. Those two effects set the $S_8$ derivatives that are only rescaled by a factor $\sqrt{0.3/\Omega_{m,\rm fid}}$ compared with $\sigma_8$ derivatives. For $w_0$ increasing (decreasing) $w_0$ decreases (increases) the variance which is opposite to the effect of $\sigma_8$ as replicated in the $\gamma-$2PCF derivatives in figure~\ref{fig:xi_derivatives}. In the noise-free case, the changes in $S_8$ and $w_0$ occur in an almost redshift independent way, we observe a redshift dependent behaviour on the noisy case due to the shape noise contribution to the variance, which is larger for the lower redshift bins. We also observe that for various redshift combinations, the $\Omega_{\rm m}$ derivatives are asymmetric with respect to this peak location, which is an indication that variations on $\Omega_{\rm m}$ affect the variance and the higher order cumulants.  

To understand the derivative behaviour, we study the cosmology dependence of the PDF that is encoded in the $\kappa-$cumulants  (see equation \eqref{eq:cumulants}). First, to gain an intuition for the changes on the cumulants with cosmology, we write equation \eqref{eq:cumulants} in terms of the reduced cumulants of the matter field in an infinitely long cylinder with transverse angular radius $R_{\theta} = \chi \theta$ as 
\begin{equation}
    \langle \kappa_{\theta}^l\rangle_{c} = \int^{\chi_s}_{0}\,d \chi\, w(\chi)^l \sigma_{\cyl}^{2(l-1)}(z,R_{\theta}) S^{2D}_l(R_{\theta})\,.
\end{equation}
We will discuss the changes in the variance $(l=2)$ and the skewness $(l=3)$. In the linear regime and a $\Lambda$CDM universe, the variance is proportional to the square of the linear growth factor $\sigma_{\cyl}^2(z, R_{\theta}) \propto D^2(z)\sigma_{\rm L}^2(R_{\theta})$, which changes with  $\Omega_{\rm m}$ and $w_0$.  

In an EDS universe the reduced skewness is affected by the scale-dependent variance through its logarithmic derivative \citep{Bernardeau:1994zd, Boyle:2020bqn} as  
\begin{align}
\label{eq:S3pred}
    S^{2D}_3(R_{\theta}) = \frac{\langle\delta^3(R_{\theta})\rangle_c}{\langle\delta^{2}(R_{\theta})\rangle_c^{2}}\approx \frac{36}{7} + \frac{3}{2}
    \frac{d\log \sigma^2_{\rm L}(R_{\theta})}{d\log R_{\theta}}\,.
\end{align}
Changes on $\Omega_{\rm m}$  of $\pm 10\%$ change the variance in a scale dependent way by $\mp 3\%$.  Note that the logarithmic derivative is redshift independent and the $\sigma_8$ dependence cancels. The variance and skewness are also affected by the information encoded in the weak lensing kernel. Hence, for $\Omega_{\rm m}$ derivatives, the redshift dependence comes from the weak lensing kernel contribution. The highest redshift bin (bin$5$) probes the largest scales (see FIG \ref{fig:weak_lens}) where the effect on the skewness is a few percent compared with the effect on the variance, therefore for this bin and combinations that include it, changes in $\Omega_{\rm m}$ mostly affect the variance and then the derivatives are more symmetric around the origin. Note, that the previous explanation holds for $\Omega_{\rm m}$ derivatives at fixed $\sigma_8$. In our case, the $\Omega_{\rm m}$ derivatives at fixed $S_8$ present a larger asymmetry due to an extra subtraction of the $\sigma_8$ derivatives that appear from the transformation \eqref{eq:chainrule}. For the non-tomographic case, we reproduce the results of figure~5 in \cite{Boyle:2020bqn} noting that their $\Omega_{\rm m}$-derivatives are at fixed $\sigma_8$. In the non-tomographic case, the integration along the line of sight combines all the scales within the redshift range, obscuring the growth of structure information. This information is recovered in the tomographic case by considering the effect in the skewness at different physical scales.
The lower redshift bins are highly affected by the shape noise. As a consequence, the effect on the skewness is erased for the lower redshift bins. The skewness contribution is still significant for a few redshift combinations that include intermediate redshift bins for which  the effect of shape noise is sufficiently small.  

\begin{figure}[h]
    \includegraphics[width=0.44\textwidth]{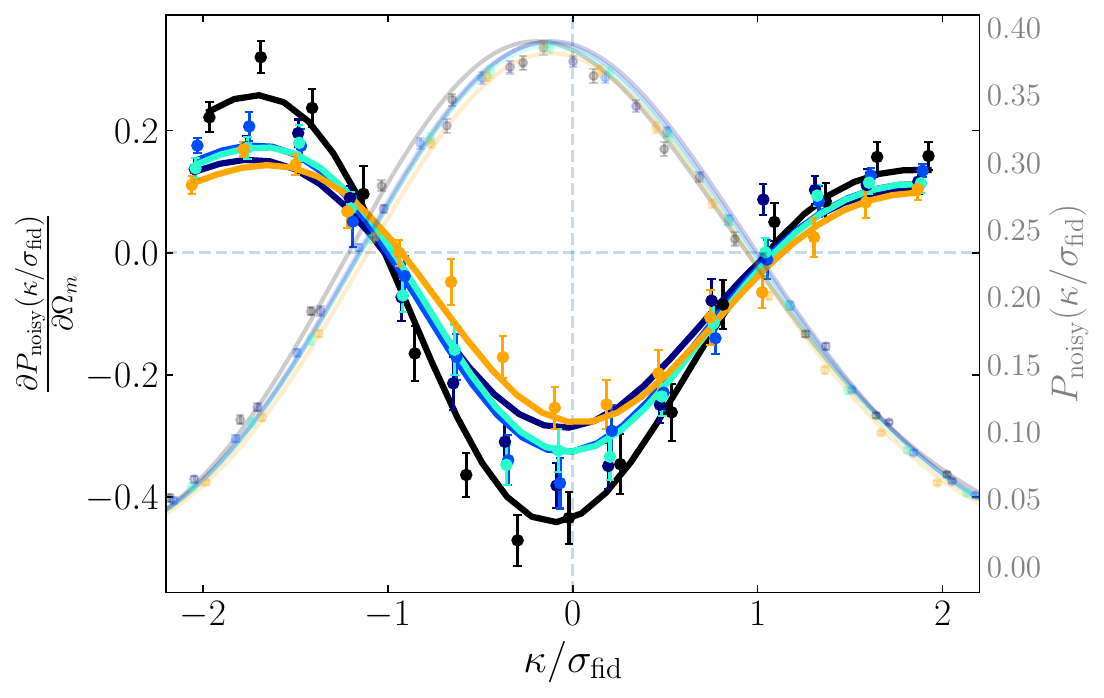} 
    \includegraphics[width=0.44\textwidth]{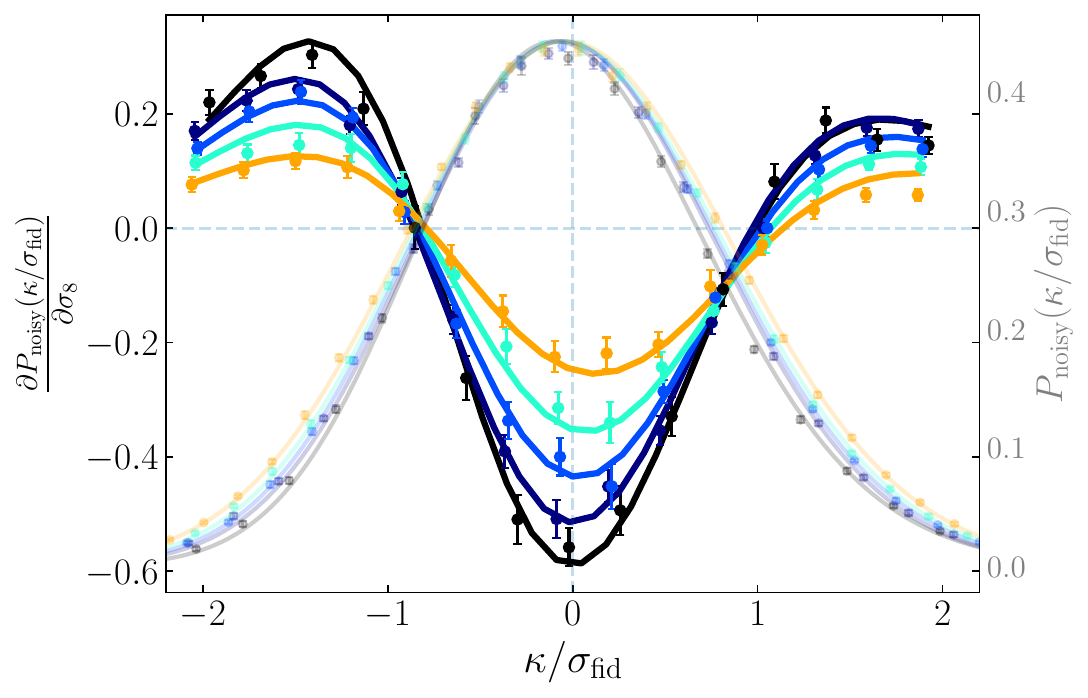} 
    \caption{Noisy PDF derivative validation for $\Omega_{\rm m}$ at fixed $\sigma_8$ (upper panel) and $\sigma_8$ (lower panel). (Solid lines) Theoretical derivative predictions for the \textit{cosmo}-SLICS cosmology. (Dots) derivative from measurements of the {\it cosmo}-SLICS $\kappa-$maps. The error bars represent the error of the mean over the $20$ {\it cosmo}-SLICS realizations. The colour scheme follows the one of FIG \ref{fig:freenoisy_vs_noisy_PDF_validation}. The light lines and dots  show the comparison between the predicted and the measured $\kappa-$PDF for the corresponding {\it cosmo}-SLICS model (See TAB. \ref{tab:Cosmo-SLICS-validation}). }
    \label{fig:derivative_validation}
\end{figure}

We validate the derivatives used in our analysis by using measurements from maps of the {\it cosmo}-SLICS catalogues. For each parameter derivative, we choose the cosmology nodes in the {\it cosmo}-SLICS hypercube for which the other parameters change the least w.r.t the {\it cosmo}-SLICS fiducial cosmology. For each cosmology node, we measure and predict the $\kappa-$PDF following the procedures presented in section \ref{sec:validation}. In table \ref{tab:Cosmo-SLICS-validation} we show which {\it cosmo}-SLICS model we use for each parameter (first column) and the corresponding values for the cosmological parameters. The derivatives are computed by taking finite differences with the fiducial {\it cosmo}-SLICS cosmology (Model fid). We present derivative validations for $\{\Omega_{\rm m},\sigma_8\}$ in figure \ref{fig:derivative_validation}, we find good agreement between the predicted and measured derivatives. For the dark energy parameter, $w_0$ we could not find a suitable validation model within {\it cosmo}-SLICS, but the theory has been previously validated in the non-tomographic case \citep{Boyle:2020bqn}. Generally, the theoretical framework for the $\kappa-$PDF including shape noise successfully recovers the cosmological dependence of \textit{cosmo}-SLICS (as well as the signal from intrinsic alignment as shown in Appendix \ref{app:IA}).
\begin{table}
\begin{centering}
\begin{tabular}{ccccc}
\hline 
 & $\Omega_{\rm m}$ & $\sigma_{8}$ & $w_{0}$ & $h$\tabularnewline
\hline 
\hline 
Model fid & $0.2905$ & $0.8364$ & $-1.0$ & $0.6898$\tabularnewline
\hline 
Model 04 $(\Omega_{m})$ & $0.3759$ & $0.8028$ & $-0.9741$ & $0.6034$\tabularnewline
\hline 
Model 18 $(\sigma_{8})$ & $0.2784$ & $0.6747$ & $-1.0673$ & $0.6747$\tabularnewline
\hline 
\end{tabular}
\par\end{centering}
\caption{\label{tab:Cosmo-SLICS-validation}{\it cosmo}-SLICS cosmologies for derivative validation.}
\end{table}

\subsection{Data vector covariance}

The covariance used for the Fisher analysis is estimated from the $\kappa-$PDF and $\gamma-$2PCF measurements from the SLICS $\kappa-$maps and shear maps respectively. The covariance is computed as
\begin{equation}
    C_{ij} =  \big\langle (D_i -\bar{D_i})(D_j-\bar{D_j})\big\rangle\,\qquad \bar{D_i} = \langle D_i \rangle,
\end{equation}
where the ensemble average is over the $N_{\rm sim}=923$ SLICS realizations, and $\{i,j\}$ run over the elements of the data vector $\bm{D}$. 

\begin{figure}[h]
\centering
    \includegraphics[width=0.46\textwidth]{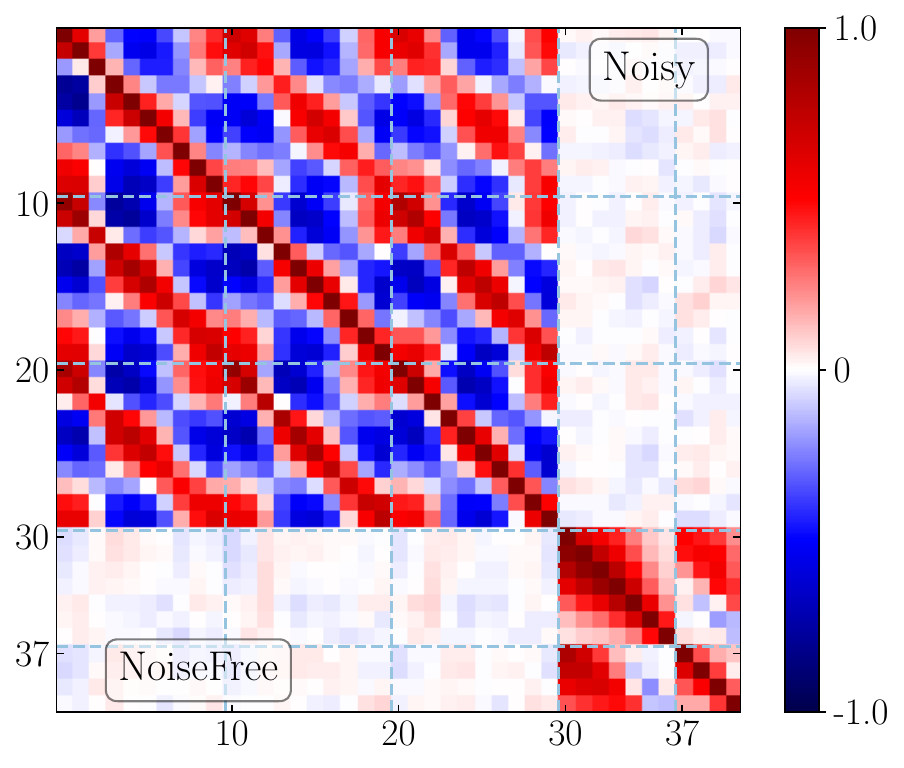} 
    \caption{Cross-correlation matrix from SLICS simulations bin$2$. $\kappa-$PDF for the smoothing scales $\theta = \{10,15,20\}$ arcmin and $\xi_{\pm}(\theta)$. The blue dashed lines indicate the boundaries between different segments of the data vector}. 
    \label{fig:r_SLICS_bin2}
\end{figure}

In order to study the correlation among the bins  in the data vector and how shape noise affect these correlations, we plot the cross correlation matrix defined as  
\begin{equation}
    r_{ij} = \frac{C_{ij}}{\sqrt{C_{ii}C_{jj}}}\,.
\end{equation}
In FIG \ref{fig:r_SLICS_bin2} we show the correlation matrix for the noise-free and noisy summary statistics, on the lower and upper diagonal triangles respectively, for the redshift bin2. The data vector in each case is the concatenation of the $\kappa-$PDF for three smoothing scales $\theta = \{10,15,20\}$ arcmin and the two components of the $\gamma-$2PCF, $\xi_{\pm}$. Each of the blocks on the $\kappa-$PDF zone contains $10$ $\kappa$ bins and shows correlations among them and correlations between different smoothing scales. We first summarise the main features of an individual $\kappa$-PDF covariance, for which a detailed explanation can be found in \cite{Uhlemann:2022znd}. Around the diagonal, consecutive $\kappa$ bins are strongly correlated due to the overlapping of the circular apertures (generated during the Top-Hat smoothing). We remove the mean convergence in our individually simulated small patches to more closely resemble the case of cutting patches from a large portion of the full sky. This makes the super-sample covariance term causing underdense and overdense bins to be negatively correlated subdominant, revealing a more complex 3x3 tiling pattern caused by the next-to-leading order term. We also observe correlations among the $\kappa-$PDF at different smoothing scales, since the circular apertures on the larger scales partially capture similar information present in the smaller ones. As expected, the cross-correlations decrease with an increasing difference in the smoothing scales. As shape noise has a stronger effect on the lower redshift bins, when comparing the lower and upper triangles we can see a slightly smaller difference for the correlations. This is less evident for the higher redshift bins as is shown in the next figure (figure \ref{fig:r_SLICS_4 }). In the presence of shape noise, the diagonal of $\kappa-$PDF covariance is mainly affected by the Top-Hat filtering which induces correlations between the shape noise values \citep{Uhlemann:2022znd}. 

The $\gamma-$2PCF zone involve correlations among the logarithmic spaced bins for each of the components $\xi_{\pm}$ with the previously discussed scale cuts. We observe strong positive correlations in $\xi_{+}$, forming a square, among small angular scales. The two components $\xi_{\pm}$ are strongly correlated too, but with a shift in the angular scales. In the presence of shape noise, all the correlations are diminished, with the correlations in $\xi_{-}$ being more affected. Additionally, we observe small correlations of $~8 \%$ between the $\kappa-$PDF and the $\gamma-2$PCF because of the correlation of the amplitude of the $\gamma-2$PCF with the $\kappa-$variance that sets the shape of the $\kappa-$PDF.      
  
The correlation matrix in FIG \ref{fig:r_SLICS_4 } consider correlations among the highest 4 auto redshift bins ($\rm bin5,\rm bin4,\rm bin3,\rm bin2$) for the $\kappa$-PDF including the 3 smoothing scales and for the two components of the $\gamma-$2PCF, $\xi_{\pm}(\theta)$. We observe in the $\kappa-$PDF zone that correlations between non-consecutive redshift bins are smaller than the ones between consecutive redshift bins. Correlations among different redshift bins are diminished because of shape noise, which affects lower redshift bins more strongly.  
Correlations among different $\xi_{+}$ measured bins are larger for the lower redshift bins, as these bins capture smaller scales compared with the highest redshift bins \citep{Joachimi:2007xd}. Correlations between the $\kappa-$PDF and the $\gamma-$2PCF range from $~ -8 \%$ to $~8\%$ across all redshift bins.     
\begin{figure}[h]
    \includegraphics[width=0.46\textwidth]{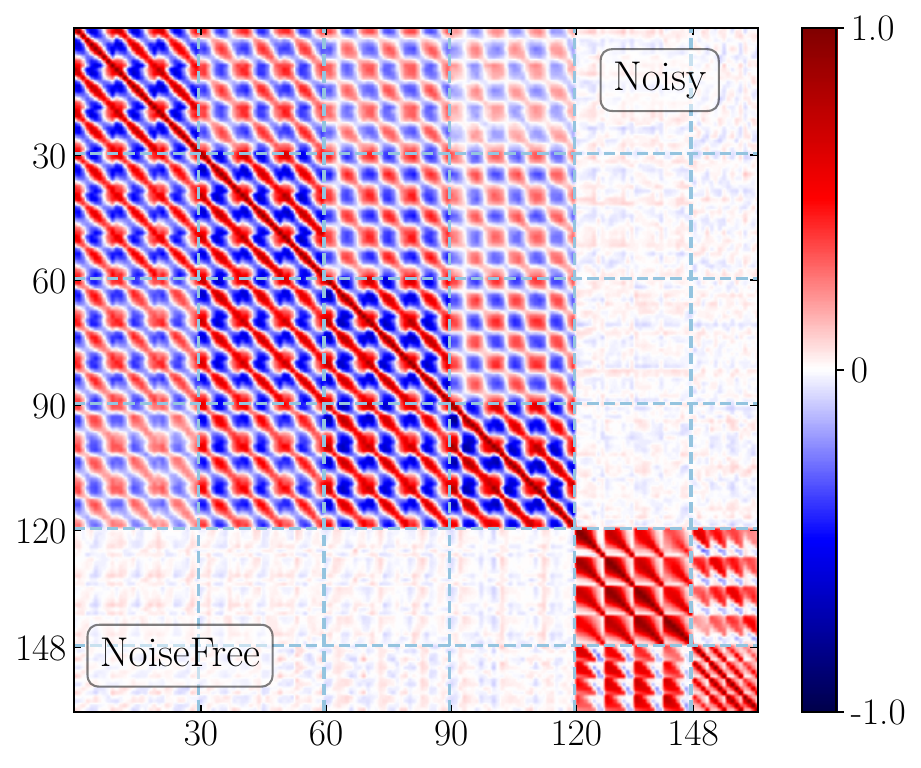} 
    \caption{Cross-correlation matrix from SLICS simulations for the case of single redshift bins and only auto-correlation functions for 4 redshift bins  ($\rm bin 5,\rm bin4,\rm bin3,\rm bin2$).}
    \label{fig:r_SLICS_4 }
\end{figure}

\subsection{Results}
\label{sec:results}
Now we present our main results from the Fisher forecast for the different cases we numbered at the beginning of this section. 

To mimic a StageIV-like survey area of $A_{\rm survey} =$15,000deg$^2$, the data covariance matrix is scaled as $A_{\rm SLICS}/A_{\rm survey}$, where $A_{\rm SLICS} =$ 100deg$^2$ is the area of the SLICS simulations. In order to mitigate the sampling noise in the covariance estimation, that may affect the parameter covariance \citep{Hartlap:2006kj, Dodelson:2013uaa},  we use the Sellentin-Heavens prescription \citep{Sellentin:2015waz} where the inverse data covariance is rescaled by the factor
\begin{equation}
    \alpha = \frac{N_{\rm sim}-N_s+N_p+1}{N_{\rm sim}-1}
    \label{eq:Sellentin-Heavens}
\end{equation}
with $N_s$ the length of the data vector $D$, $N_{\rm sim}$ is the number of realisations for the covariance estimate and $N_p$ is the number of parameters to estimate.

The predicted marginalised 1-$\sigma$ uncertainties on the set of parameters $p_i = \{\Omega_{\rm m},S_8,w_0\}$ are determined from the Fisher matrix from equation~\eqref{eq:Fisher_matrix}, as
\begin{equation}
   \Delta p_i = \sqrt{(F^{-1})_{ii}}\,.
   \label{eq:eq:uncertainty}
\end{equation}
We summarise the forecasted parameter uncertainties for the different cases in Table \ref{tab:Fisher_Uncertainties}. 
\begin{table*}
\centering
\begin{tabular}{lccc}
\toprule 
 & $\Delta\Omega_{m}\left(10^{-3}\right)$ & $\Delta S_{8}\left(10^{-3}\right)$ & $\Delta w_{0}\left(10^{-2}\right)$\tabularnewline
\midrule
\midrule 
$P(\kappa)$ No Tomography  & $9.79$ & $3.29$ & $3.85$\tabularnewline
\midrule 
$P(\kappa)$ Tomography  & $1.88$ & $2.11$ & $1.65$\tabularnewline
\midrule 
$P_{{\rm noisy}}(\kappa)$ No Tomography  & $10.74$ & $3.92$ & $4.36$\tabularnewline
\midrule 
$P_{{\rm noisy}}(\kappa)$ Tomography only auto & $4.32$  & $3.53$ & $3.19$\tabularnewline
\midrule 
$P_{{\rm noisy}}(\kappa)$ Tomography  & $2.79$ & $2.99$ & $2.65$\tabularnewline
\midrule 
$\xi(\theta)$ Tomography  & $3.47$ & $4.85$  & $3.40$\tabularnewline
\midrule 
$P_{{\rm noisy}}(\kappa)$ + $\xi$ Tomography & $2.08$ & $2.72$ & $2.34$\tabularnewline
\bottomrule
\end{tabular}
\caption{Fisher forecast. $1\sigma$ predicted uncertainties (equation \ref{eq:eq:uncertainty}) for the set of parameters $\{\Omega_{\rm m},S_8,w_0\}$.}
\label{tab:Fisher_Uncertainties}
\end{table*}

In the following, we investigate the constraining power of the tomographic analysis with the $\kappa-$PDF and the $\gamma-$2PCF. We start by studying the effect of tomography  on the constraints compared with the non-tomographic case.

\begin{figure}[h]
     \includegraphics[width=0.45\textwidth]{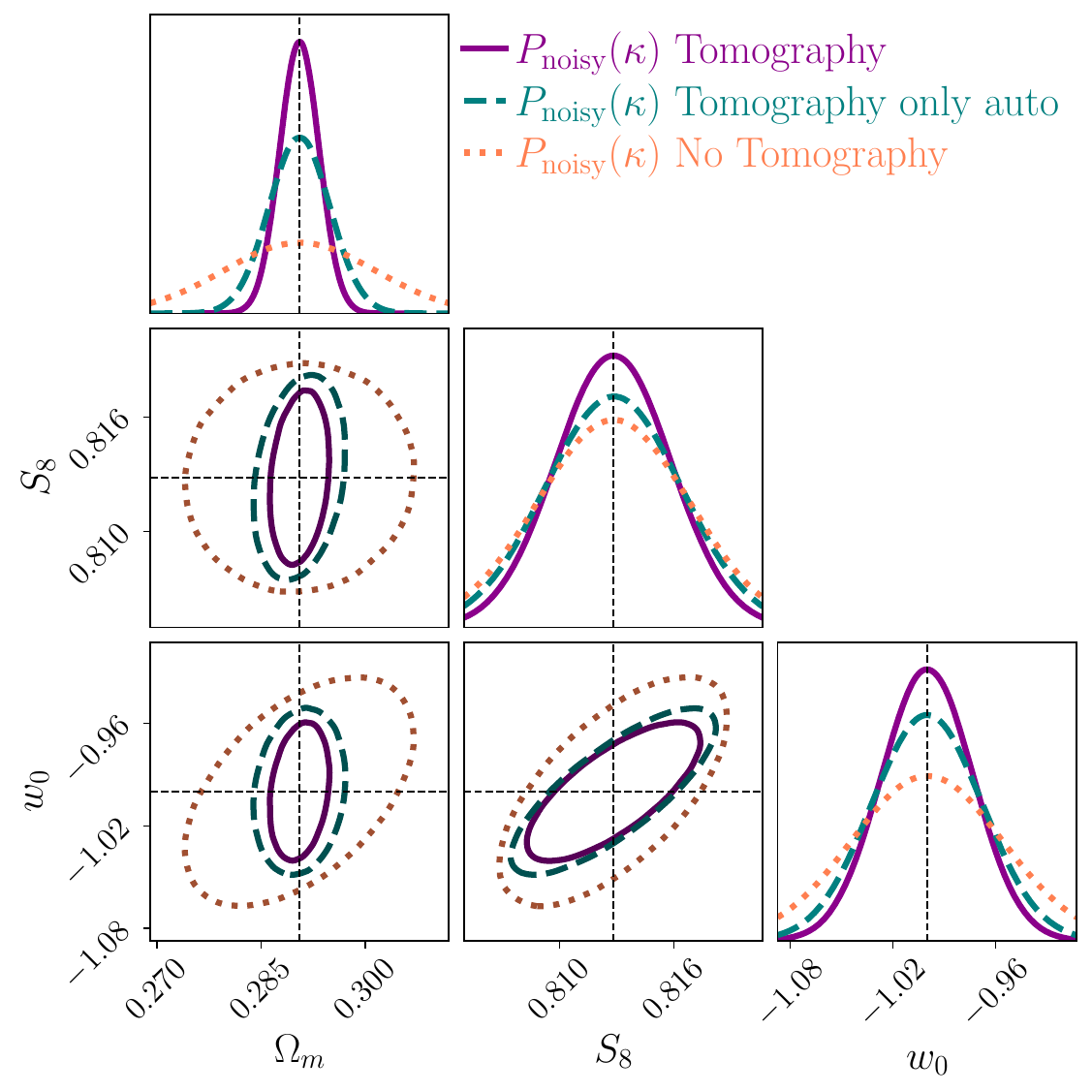} 
    \caption{$1\sigma$ Fisher forecast constraints for $\{\Omega_{\rm m},S_8,w_0\}$ from the $\kappa-$PDF. Tomography including 14 redshift combinations (solid violet contour). Tomography including only single and ``All bins" redshift bins (dashed teal contour). No tomography (only ``All bins", dotted coral contour). In all cases we combine three smoothing scales $\theta=\{10,15,20\}$ arcmin.} 
    \label{fig:Fisher_tomo_no_tomo}
\end{figure}

\paragraph{Tomographic enhancement of the $\kappa-$PDF constraints}
We compare the constraints obtained by performing a tomographic and non-tomographic analysis with the $\kappa-$PDF. In FIG \ref{fig:Fisher_tomo_no_tomo} we show Fisher contours obtained with the noisy $\kappa-$PDF for: the non-tomography case, tomography including the 4 higher individual redshift bins along with the non-tomographic case, and the tomography with 14 redshift bin combinations. We clearly see how the  tomography tightens constraints. In all cases we combine three smoothing scales $\theta = \{10,15,20\} $ arcmin.  As illustrated, in the presence of shape noise, the  constraints on $\{\Omega_{\rm m},S_8,w_0\}$ are improved by a factor of \{3.8,1.3,1.6\}. For the noise-free $\kappa-$PDF the constraints are improved by a factor of $\{5.2,1.6,2.3\}$. The significant enhancement of the constraints on $\Omega_{\rm m}$ is due to the additional information provided for the skewness and the higher order moments at different redshift bins,  as we explained in section \ref{sec:derivatives}. By considering different redshift bin contributions, it is possible to recover the information of the growth of structures at different scales. Hence, tomography allows disentangling $\Omega_{\rm m}$ from $w_0$, by distinguishing the growth effect of $w_0$ from the combined growth and skewness effect in $\Omega_{\rm m}$.

The  noise-free and noisy $\kappa-$PDF constraints are degraded by approximately a factor of $1.5$ for all parameters when adding shape noise. Degradation in the constraints is caused by a reduction in the amplitude of the noisy PDF derivatives of about $30\%$ as shown in figure~\ref{fig:PDF_derivatives}, with the lower redshifts combinations being the more affected, and because the shape noise increases the diagonal of the data covariance \citep{Uhlemann:2022znd}. Those two effects reduce the signal-to-noise only degrading  the constraints, keeping the orientation of the contours almost unchanged.

\begin{figure}[h]
    \includegraphics[width=0.45\textwidth]{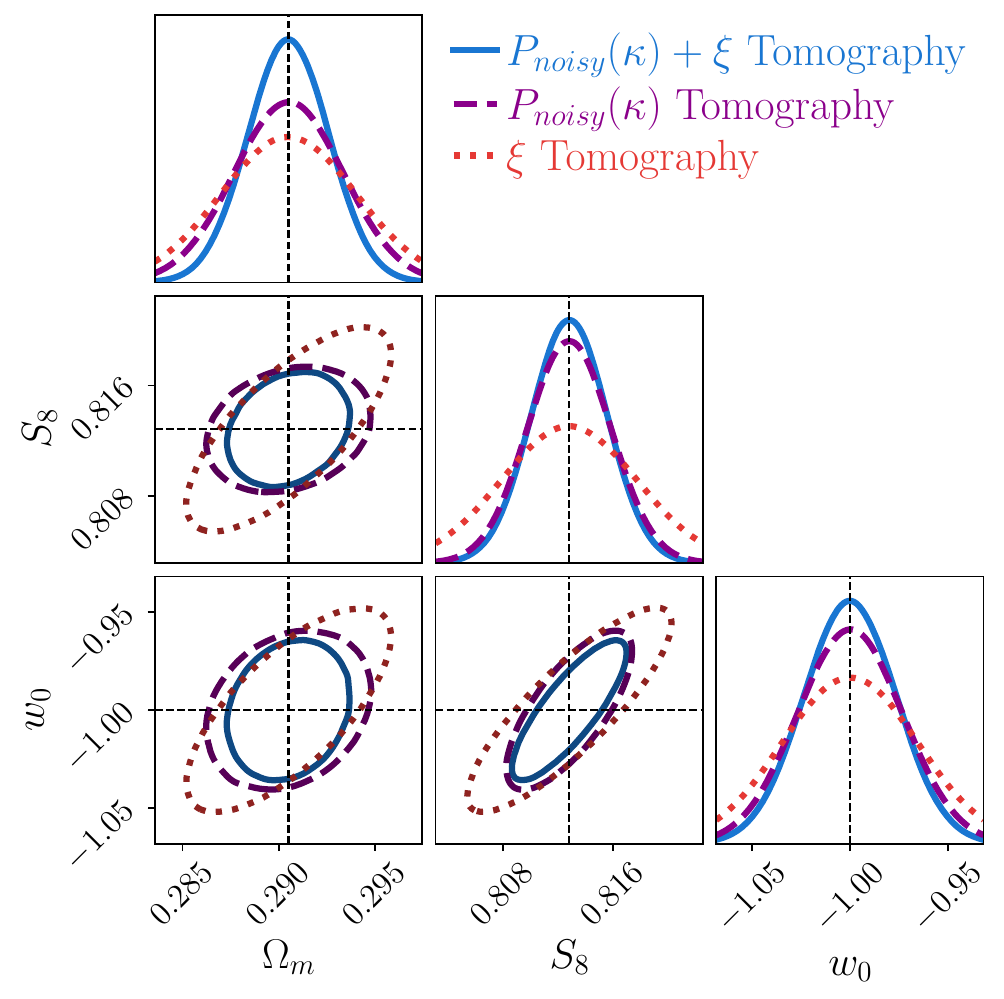} 
    \caption{Joint analysis: $1\sigma$ Fisher forecast constraints from the $\kappa-$PDF and $\gamma-$2PCF. Noisy $\kappa-$PDF (dashed violet contour) combining three smoothing scales $\theta=\{10,15,20\}$ arcmin. $\xi = \xi_{+} + \xi_{-}$ (dotted red contour), for scales $\theta>~10$ arcmin and $\theta>~50$ arcmin for $\xi_{\pm}$ respectively. Combined $\kappa-$PDF and $\xi-$2PCF are shown with the solid blue contours.} 
    \label{fig:fisher_joint_pdf_2pcf}
\end{figure}

\paragraph{Joint analysis $\kappa-$PDF and $\gamma-$2PCF}
In figure \ref{fig:fisher_joint_pdf_2pcf} we present our main result where we combine the $\kappa-$PDF and $\gamma-$2PCF, including both components $\xi_{\pm}$. We include the $\kappa-$PDF for the three smoothing scales $\{10,15,20\}$ arcmin (see appendix \ref{app:joint} for a study with a single smoothing scale) and consider scales $\theta > 10$ arcmin and $\theta >50$ arcmin for $\xi_{\pm}$, respectively (see the predicted $\xi_{\pm}$ derivatives in  FIG \ref{fig:xi_derivatives}). The $\kappa-$PDF outperforms the $\gamma-$2PCF by approximately  $23\%$ for $\Omega_{\rm m}$, $57\%$ for $S_8$ and a $53\%$ for $w_0$. The improvement stems from the non-Gaussian information encoded in the skewness and higher order moments within the $\kappa-$PDF which breaks the degeneracy between $w_0$ and $\sigma_8$ that is pronounced for the $\gamma-$2PCF as illustrated in figure~\ref{fig:xi_derivatives}. Combining both summary statistics leads to an improvement in the constraints by a factor of  $\{1.7,1.8,1.5\}$ for $\{\Omega_{\rm m},S_8,w_0\}$ respectively compared with the $\gamma-$2PCF. 

\begin{figure}[h]
    \includegraphics[width=0.45\textwidth]{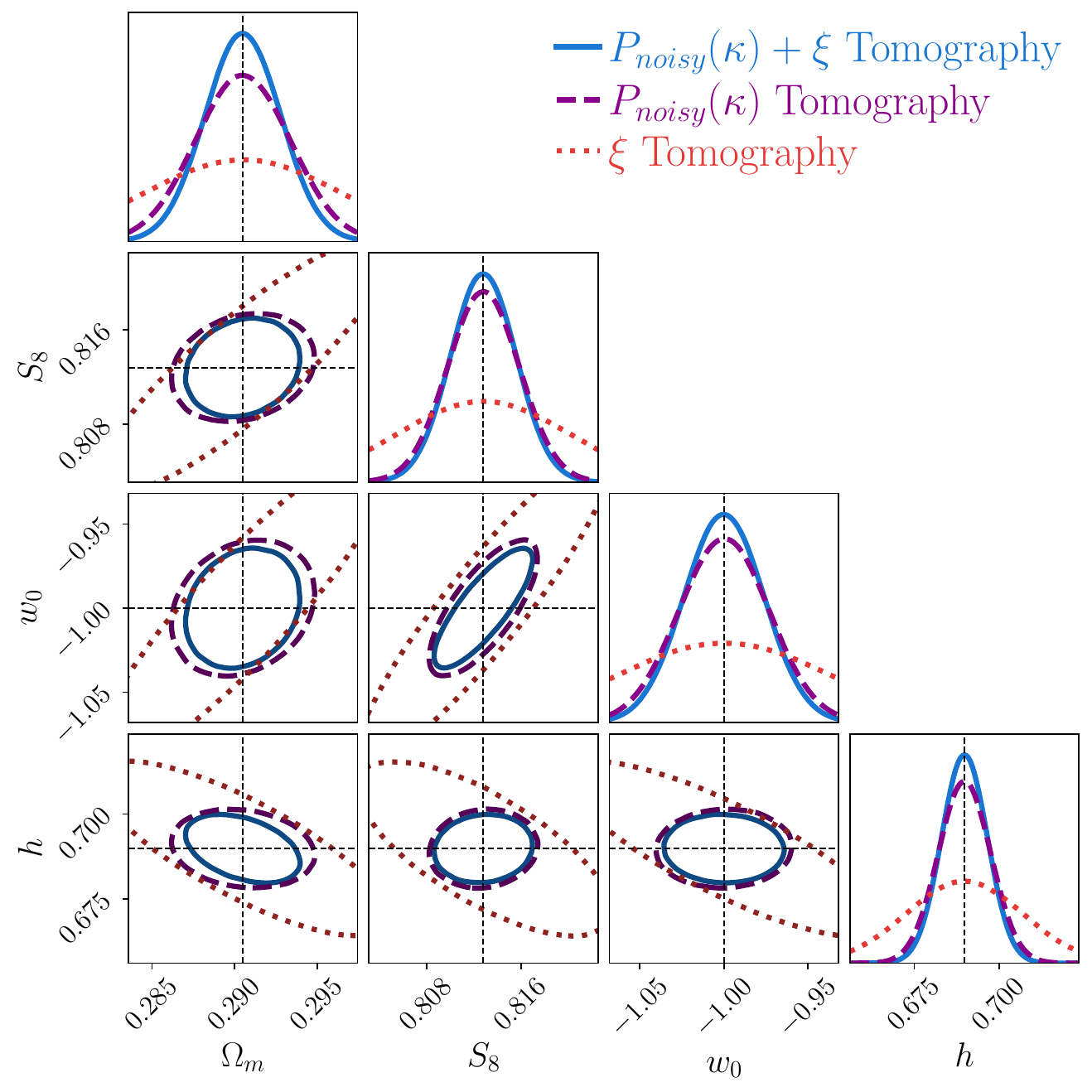} 
    \caption{Fisher forecast for $1-\sigma$ constraints from the joint $\gamma-$2PCF and $\kappa$-PDF tomography as in figure~\ref{fig:fisher_joint_pdf_2pcf} including $h$.} 
    \label{fig:fisher_joint_pdf_2pcf_with_h}
\end{figure}

We have observed that the tomographic $\kappa-$PDF holds strong potential for constraining the parameter set $\{\Omega_{\rm m},S_8,w_0\}$. This is because it can track the growth of structure and capture non-Gaussian information across different scales. We now investigate how these conclusions are affected when forecasting for a different number of parameters. We begin by including the Hubble parameter, considering $~10\%$ steps around the fiducial value, $h^{-}=0.6198 $, $h^{+} = 0.7598$. Changing $h$ primarily affects the variance in a redshift-dependent manner and impacts the skewness without modifying the growth factor like $\Omega_{\rm m}$. Both $\Omega_{\rm m}$ and $h$ change the variance and skewness across the redshift bins, while $S_8$ and $w_0$ do not (significantly). Therefore, constraints with varying $h$ benefit significantly from the additional non-Gaussian information at different scales, leading to improved constraints as shown in figure  \ref{fig:fisher_joint_pdf_2pcf_with_h}. In this case, the joint analysis with the $\kappa-$PDF and the $\gamma-$2PCF offers a constraint improvement of about a factor of 2.5 for $\{\Omega_{\rm m},S_8,w_0,h\}$ compared with the $\gamma-$2PCF alone. This is because the constraints from the $\gamma-$2PCF are degraded by a factor of $\{1.7,1.4,1.8\}$  for $\{\Omega_{\rm m},S_8,w_0\}$ respectively when adding $h$, while the constraints from $\kappa-$PDF only degrade by less than  $2\%$.

\begin{figure}[h]
    \includegraphics[width=0.45\textwidth]{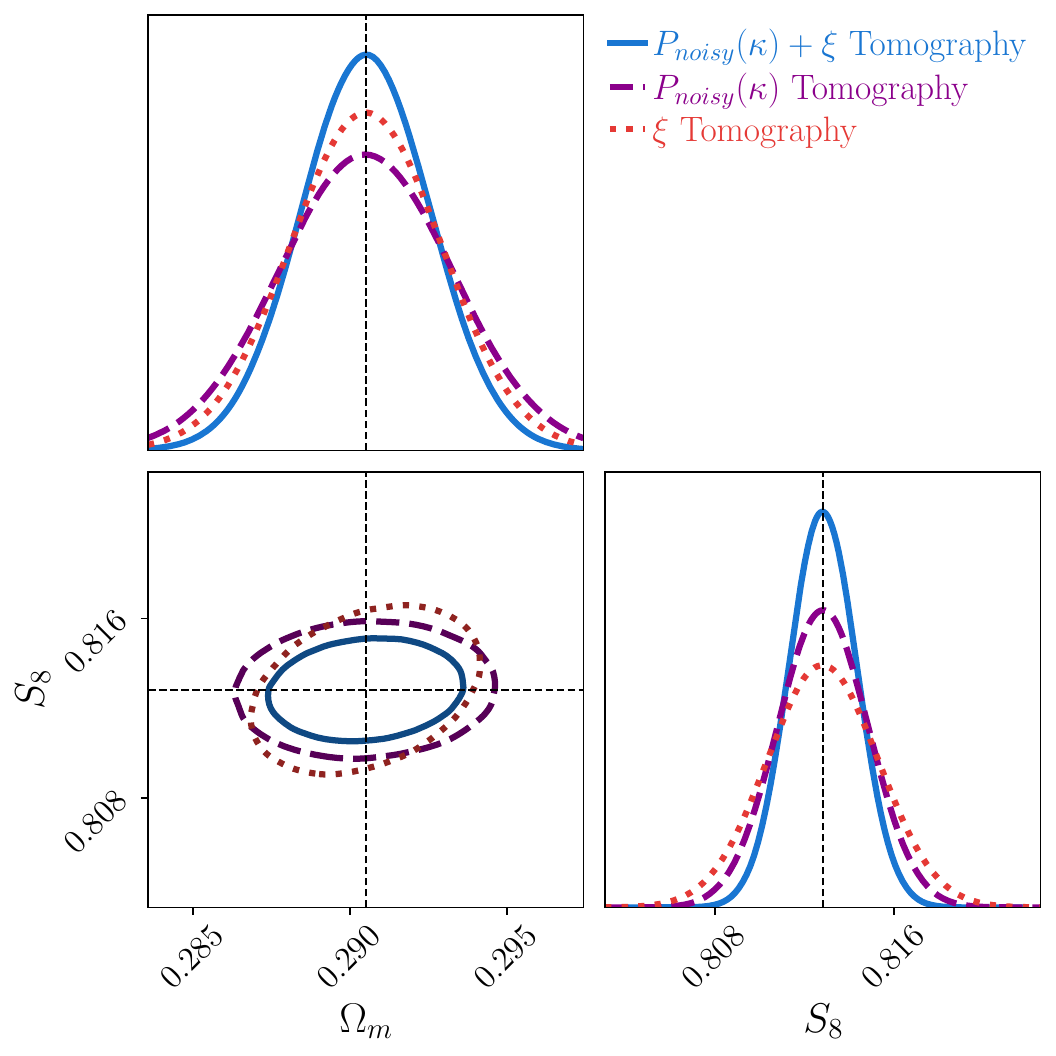} 
    \caption{Fisher forecast for $1-\sigma$ constraints from the joint $\gamma-$2PCF and $\kappa$-PDF tomography as in figure~\ref{fig:fisher_joint_pdf_2pcf} but for fixed ($w_0$, $h$).} 
    \label{fig:fisher_joint_pdf_2pcf_now}
\end{figure}

We now focus on forecasting just two parameters, holding the other two fixed. In figure \ref{fig:fisher_joint_pdf_2pcf_now} we show the $1-\sigma$ fisher contours for $\Omega_{\rm m}$ and $S_8$ at fixed $w_0$ (for other two-parameter cases, see figure \ref{fig:fisher_joint_pdf_2pcf_nox} in Appendix~\ref{app:joint}). When constraining only two cosmological parameters, the $\kappa-$PDF and the $\gamma-$2PCF provide similar constraints with slightly different degeneracy directions between the parameters. This helps to improve the constraint when combining both summary statistics. Hence, when considering more than two parameters, the $\gamma-$2PCF loses constraining power due to a degeneracy between the parameters, different from the $\kappa-$PDF that is able to disentangle all  parameters obtaining better constraints.

\paragraph{Relation to previous works} Our findings regarding the strong constraining power of the tomographic $\kappa-$PDF across mildly nonlinear scales and its complementarity with the $\gamma-$2PCF are consistent with  previous works. For example, the non-tomographic PDF from \cite{Boyle:2020bqn}, which is based on the LDT, demonstrated the power of the $\kappa-$PDF when combining different smoothing scales and the complementarity for 2PCF and CMB data, in particular for dark energy. We find that tomography not only preserves but enhances this information. \cite{Thiele:2020rig} employed a halo-model formalism, originally formulated for the thermal Sunyaev-Zel’dovich effect, to compute the one and two-point convergence PDFs, thus including the PDF covariance. This analytical model accurately describes the positive tail of the PDF describing large convergence values dominated by massive halos. However, it is strongly sensitive to small-scale effects and is less accurate on mildly nonlinear scales and negative convergence values. The authors perform a Fisher forecast for key cosmological parameters like the total neutrino mass, matter density, and the amplitude of density fluctuations. Their analysis showed that the convergence PDF matches or outperforms constraints from the power spectrum  and that by combining both statistics constraints improve. \cite{Liu2019} performed a tomographic analysis that combines the convergence PDF and power spectrum to constrain neutrino masses. As the impact of neutrinos is stronger on small scales the authors use  cosmological massive neutrino simulations that provides PDF measurements valid in the one-halo regime. They found that tomography has the potential to significantly improve the constraints, obtaining tighter constraints with the PDF than with the power spectrum. Finally, \cite{Giblin:2022ucn} used $\kappa-$PDFs measurements from the {\it cosmo}-SLICS simulations to train a Gaussian process regression emulator and carry out an MCMC analysis to constrain $\{\Omega_{\rm m},S_8,h,w_0\}$. The authors demonstrate the constraining power of the tomographic convergence PDF at three highly non-linear scales (2,7 and 13 arcmin) and demonstrate that combining different smoothing scales leads to  improve the cosmological parameter constraints.

\section{Conclusion}\label{sec:conclusions}

\paragraph{Summary} In this paper, we have studied the constraining power of the tomographic one-point PDF for the weak lensing convergence field. We consider a StageIV-like galaxy distribution which we have sliced into $5$ equipopulated redshift bins. We predict the $\kappa-$PDF using the public \cosmomentum code \citep{Friedrich:2019byw} based on the large deviation theory (LDT). We reconstruct the convergence field from the shear components by performing a KS inversion and forward model its impact on the non-linear variance following the procedure of \citet{Barthelemy:2023mer}. To account for shape noise in $\kappa-$PDF predictions, we convolve with a Gaussian (proportional to intrinsic ellipticity variance). Since  shape noise impacts the lower redshift bins more, we propose a tomographic strategy by considering the combinations of consecutive redshift bins (quarters, triples, pairs) to minimise noise. 

We validated the theoretical predictions with measurements from numerical cosmic shear simulations sampling multiple $w$CDM cosmologies. We found good agreement between the predictions and measurements, with  errors below $5\%$ in the 2-sigma around the peak of the PDF for both the noise-free and noisy cases for all redshift bins. In the presence of shape noise, the predictions become more accurate as the  Gaussian shape noise conceals nonlinear effects from small physical scales at low redshifts. For comparison, we also measured $\gamma-$2PCF from the SLICS shear maps and validated them with \pyccl predictions. 

We have performed a Fisher forecast to constrain $\{\Omega_{\rm m}, S_8,w_0\}$.  We compute the derivatives from finite differences of the theoretical predictions and estimate the covariance from measurements on the fiducial SLICS maps. The tomographic analysis with the $\kappa-$PDF enhanced the constraining power compared to the non-tomographic case. In particular, there is a significant enhancement on the $\Omega_{\rm m}$ constraints due to the additional information on the growth of structure and the skewness at different redshift bins which allows disentangling $\Omega_{\rm m}$ from $w_0$. As expected, the $\kappa-$PDF outperforms the $\gamma-$2PCF, especially on $S_8$ and $w_0$ by capturing additional non-Gaussian information from higher order moments. A joint analysis of the $\kappa$-PDF and the $\gamma-$2PCF improves constraints by around a factor of $1.5$ compared with the $\gamma-$2PCF alone. 

In summary, the tomographic weak lensing PDF is a powerful tool to extract new information from mildly nonlinear scales. The $\kappa-$PDF is not only easy to measure, it is also accurately predicted on mildly non-linear scales. It shows strong potential in constraining cosmological parameters by accessing the non-Gaussian and growth structure information across different redshift bins. Thus, moving beyond two-point correlation functions will be the key to increase the precision in the inference of cosmological parameters in the era of Stage-IV surveys. 

\paragraph{Outlook.} For measuring the $\kappa-$PDF in a realistic survey mask, we can select a coverage threshold above which smoothed cells are included. For stringent thresholds close to 100\% as suggested in \cite{Barthelemy:2023mer} considering only observed pixels is sufficient. For smaller coverage thresholds around $80\%$, we could
apply the same approach as used for density-split statistics \citep{Gruen_2018} to ensure an equal masking fraction across all cells. 
Our theoretical model for the tomographic $\kappa$-PDF can be adapted to include further weak lensing systematics as discussed in \cite{Barthelemy:2023mer}. In particular, intrinsic alignments can be included for the nonlinear-alignment model, which effectively modifies the weight function by adding a term proportional to the source redshift distribution. In Appendix~\ref{app:IA} we study and include this effect in our model for the $\kappa-$PDF and show that it reproduces the \textit{cosmo}-SLICS measurements including IA with excellent accuracy. We will use this model for upcoming data analyses in future work. Additional systematic effects can be incorporated into the analysis. Photometric redshift errors affecting the $n(z)$ and hence the lensing weight could be propagated through the theoretical pipeline. Multiplicative and additive shear biases can be added, and perturbative recipes have indicated that the $\kappa-$PDF can help break degeneracies present in the $\gamma-$2PCF \citep{Patton2017}. On our mildly nonlinear scales, baryonic effects can likely be captured through their impact on the variance \citep[as also adopted in the HSC Y1 analysis][]{thiele2023hsc,Barthelemy:2023mer,Grandon:2024pek}.  
The simulated covariances can be replaced by mock catalogues obtained from shifted log-normal fields that have proven highly successful in covariance modelling for density-split statistics \citep{DES:2017hhj} and the $\kappa$-PDF \citep{Boyle:2022msq,Uhlemann:2022znd}, for which analytical recipes are available. The tomographic $\kappa$-PDF could be enhanced through an extension to joint PDFs of multiple redshift bins, most efficiently captured through a nulling procedure that localises the weighting kernels thus simplifying the correlation structure \citep{Barthelemy2022jointnulled}. This localisation might also allow to probe further into the tails of the distribution, provided that systematics can be controlled or modelled.

\section*{Acknowledgements}
LC, CU and JHD acknowledge funding from the STFC Astronomy Theory Consolidated Grant ST/W001020/1 from UK Research \& Innovation. CU's research was in part funded by the European Union (ERC StG, LSS\_BeyondAverage, 101075919). 
JHD is supported by an STFC Ernest Rutherford Fellowship (project reference ST/S004858/1). AB's work is supported by the ORIGINS excellence cluster.
The authors thank the members of the Euclid work package for higher-order weak lensing statistics and the Rubin LSST:DESC higher-order statistics topical team for useful discussions. CU thanks the \href{https://gccl-rub.github.io/}{German Centre for Cosmological Lensing} for the  integration into the national lensing community. CU is grateful for the hospitality of Perimeter Institute where the finalisation of this work following peer review was carried out. Research
at Perimeter Institute is supported in part by the Government of Canada through the Department of
Innovation, Science and Economic Development and by the Province of Ontario through the Ministry of
Colleges and Universities. This research was also supported in part by the Simons Foundation through the
Simons Foundation Emmy Noether Fellows Program at Perimeter Institute.

\appendix

\section{CGF of density in cylinders from Large deviation theory} \label{app:CGF}

The large deviation theory (LDT) is a mathematical method to quantify the probability of observing rare events on the tails of a probability distribution  that deviates from typical values \citep{dembo2009large}. As the large scale structure of the universe is non-Gaussian on intermediate and small scales, the LDT allows quantifying \textit{the asymptotic exponential shape} of the PDF for the density field $\rho$ as the variance $\sigma^2$ goes to zero \citep{Bernardeau:2015khs}. The PDF satisfy the large deviation principle if the rate function $\Psi_\rho(\rho)$, that characterises the exponential decay of the PDF, satisfy the following condition 
\begin{equation}
    \Psi_{\rho}(\rho) =-\lim_{\sigma^2\xrightarrow{}0} \sigma^2 \log\left(P(\rho)\right)\,. 
\end{equation}
 
The LDT uses the contraction principle, which gives the most likely map between the initial and final density. For line-of-sight projected fields, the initial conditions have  cylindrical symmetry, hence, the most likely mapping between the initial and final density contrast is the cylindrical collapse \citep{Valageas:2001zr}. For Gaussian initial conditions, the rate function for the linear density contrast $\delta_{\rm L}$ in cylinders is then given by
\begin{equation}
    \Psi_{\delta_{\rm L}}(\delta_{\rm L}) = \frac{\sigma_{\rm L}^2(\chi\theta,z)}{2\sigma_{\rm L}^2(r_{\rm ini},z)}\delta_{\rm L}^2\,,
\end{equation}
where $\sigma_{\rm L}^2$ is the linear cylindrical variance in the Lagrangian radius $r_{\rm ini}$. $r_{\rm ini}$ is related to the dimension and surface density $\rho$ of the cylinder through mass conservation (assuming long cylinders) as
\begin{equation}
    r_{\rm ini} = \chi\theta \rho^{1/2}\,.
\end{equation}

The time evolution of the  initial fluctuation $\zeta_{\rm SC}(\delta_{\rm L})$ and the linear density contrast $\delta_{\rm L}$ is driven by the following differential equation \citep{Bernardeau:1992zw,Bernardeau:1994hn}
\begin{equation}
    \zeta_{\rm SC}\delta_{\rm L}^2 \zeta_{\rm SC}'' - c \left(\delta_{\rm L}\zeta_{\rm SC}'\right)^2 + \frac{3}{2} \zeta_{\rm SC} \zeta_{\rm SC}' - \frac{3}{2}\zeta_{\rm SC}^2(\zeta_{\rm SC}-1) = 0\,,
\end{equation}
where the primes stand for derivatives with respect to $\delta_{\rm L}$. $c_{2D} = \frac{3}{2}$ and $c_{3D} = \frac{4}{3}$ for $2D$ and $3D$, respectively. In an EDS background, the solution can be written as  
\begin{equation}
    \zeta(\delta_{\rm L}) = \left(1-\frac{\delta_{\rm L}}{\nu}\right)^{-\nu}\,, \qquad \delta_{\rm L} = \nu \left(1-\rho^{-\frac{1}{\nu}}\right)\,,
\end{equation}
this is a fitting function that respects asymptotic behaviour in the $\delta_{\rm L}$ goes to $- \infty$ regime. The value of $\nu$ depends on the dimension and is set to reproduce the tree-order reduced skewness $S_3$, which captures the behaviour for small and intermediate values of $\delta_{\rm L}$. For cylindrical symmetry and considering infinitely long cylinders the skewness in an EDS universe is given by \citep{Uhlemann:2017kvh} 
\begin{align}
    S_3^{\rm 2D}(R) = \frac{\langle\delta^3(R)\rangle_c}{\langle\delta^2(R)\rangle_c^2} =\frac{36}{7} + \frac{3}{2}
    \frac{d\log \sigma^2_{\rm L}(R)}{d\log R}\,,
\end{align}
such that $\nu_{2D} = 1.4$.

Within the LDT theory, the scaled cumulant generating function (SCGF)\footnote{The CGF and the reduced cumulant generating function SCGF are defined as 
$$ \phi_{X}(\lambda)=\sum_{n=1}^{\infty}\langle \delta^n\rangle\frac{\lambda^n}{n!}\,,\quad \varphi_{X}(\lambda) = \lim_{\langle \delta^2\rangle\rightarrow 0} \sum_{n=0}^{\infty} S_n\frac{\lambda^n}{n!}\,,\quad S_n = \frac{\langle \delta^n\rangle}{\langle \delta^2\rangle^{n-1}}\,,$$} is predicted from Varadhan’s theorem and is given by the Legendre-Fenchel transformation of the rate function $\Psi_{\rho}(\rho)$ as  
\begin{equation}
    \varphi (\lambda) = \sup_{\rho} \left[\lambda \rho -\Psi_{\rho}(\rho)\right]\,,
\end{equation}
or as a simple Legendre transformation if $\Psi_{\rho}(\rho)$ is convex 
\begin{equation}
    \varphi (\lambda) = \lambda \rho -\Psi_{\rho}(\rho)\,,\qquad \lambda = \frac{\partial\Psi_{\rho}}{\partial\rho}\,.
\end{equation}
The SCGF of the 3D density contrast in a cylindrical filter of transverse size $\chi\theta$ and long length $L$ is given at leading order by 
\begin{equation}
    \varphi^{\text{l.o}}_{\cyl}(\lambda) = \sup_{\rho} \left[\lambda\delta_{\cyl}-\frac{\sigma_{\rm L}^2(\chi\theta,L,z)}{2\sigma_{\rm L}^2(\chi\theta(1+\delta)^{1/2},L,z)}\delta_{\rm L}^2(1+\delta)\right].
\end{equation}
The SCGF and the CGF are related in the zero variance limit as 
 \begin{equation}
    \varphi_{\cyl}^{\rm l.o}(\lambda) = \sigma^2_L \phi^{\rm l.o}\left(\frac{\lambda}{\sigma_{\rm L}^2}\right), \quad \phi_{\cyl}^{\rm l.o}(\lambda) = \frac{1}{\sigma^2_L} \varphi^{\rm l.o}\left(\lambda \sigma_{\rm L}^2\right).
    \label{eq:CGF_SCGF}
 \end{equation}
The LDT formalism gives the cumulants of the non-linear density contrast at the leading order in perturbation theory. As a consequence, the variance given by the CGF is not valid in the mildly non-linear regime and beyond. It is convenient to rescale the CGF by the non-linear variance $\sigma_{\cyl}^2 = \langle \delta_{\chi\theta,\cyl}^2\rangle$ such that the rate function on cylinder becomes
\begin{equation}
    \Psi_{\cyl}(\rho) = \frac{\sigma_{\rm L}^2(\chi\theta,z)}{\sigma_{\cyl}^2(\chi\theta,z)} \frac{\delta_{\rm L}^2(\rho)}{2\sigma_{\rm L}^2(\chi\theta\rho^{1/2},z)}\,.
\end{equation}
and then the following equations \eqref{eq:CGF_SCGF} the CGF is given by
\begin{equation}
\label{eq:rescaledCGF}
   \phi(\lambda)_{\cyl} = \frac{\sigma_{\rm L}^2}{\sigma_{\cyl}^2} \phi^{\text{l.o}}_\cyl\left(\frac{\sigma_\cyl^2}{\sigma_{\rm L}^2}\lambda\right)\,
\end{equation}
The non-linear variance can be obtained from prediction of the power spectrum, for example from \halofit  \citep{Smith_2003_halofit,peacock2014halofit} or \hmcode  \citep{Mead:2016zqy,Mead:2020vgs}.

\section{Gaussianity test for $\kappa-$PDF bins}
\begin{figure}[h]
    \includegraphics[width=0.45\textwidth]{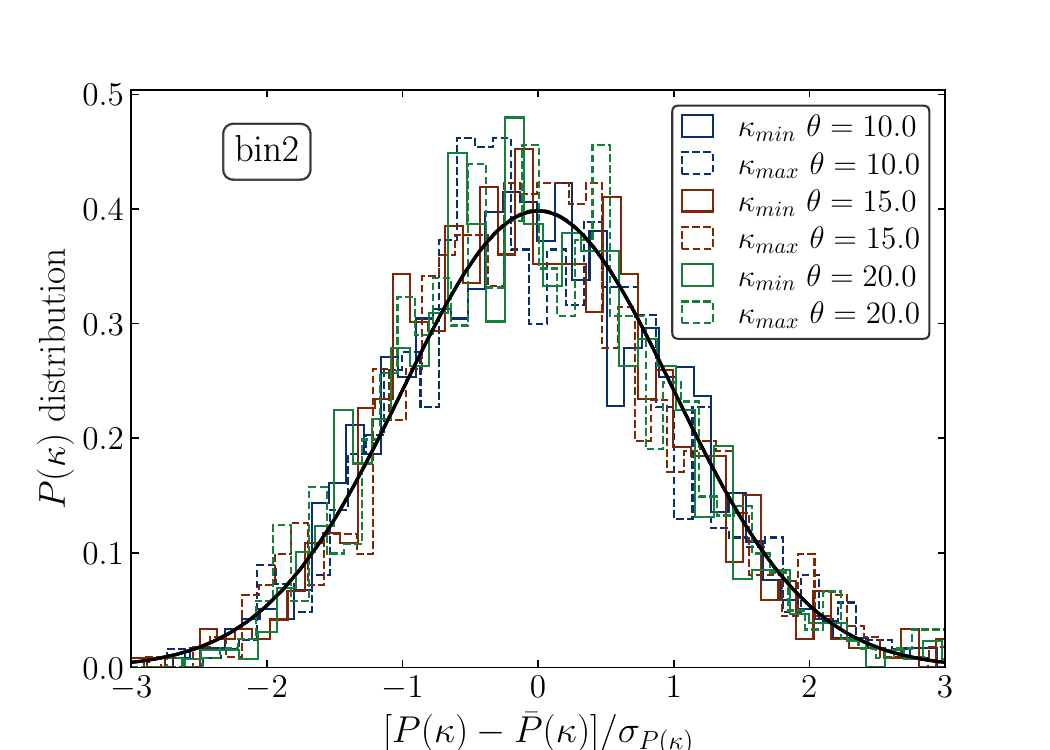} 
    \caption{Distribution of realisations around the mean for bin2 including three smoothing scales. Solid black like represent the Gaussian distribution with mean zero and unit variance.} 
    \label{fig:Gaussian_test}
\end{figure}

\begin{figure}[h]
    \includegraphics[width=0.45\textwidth]{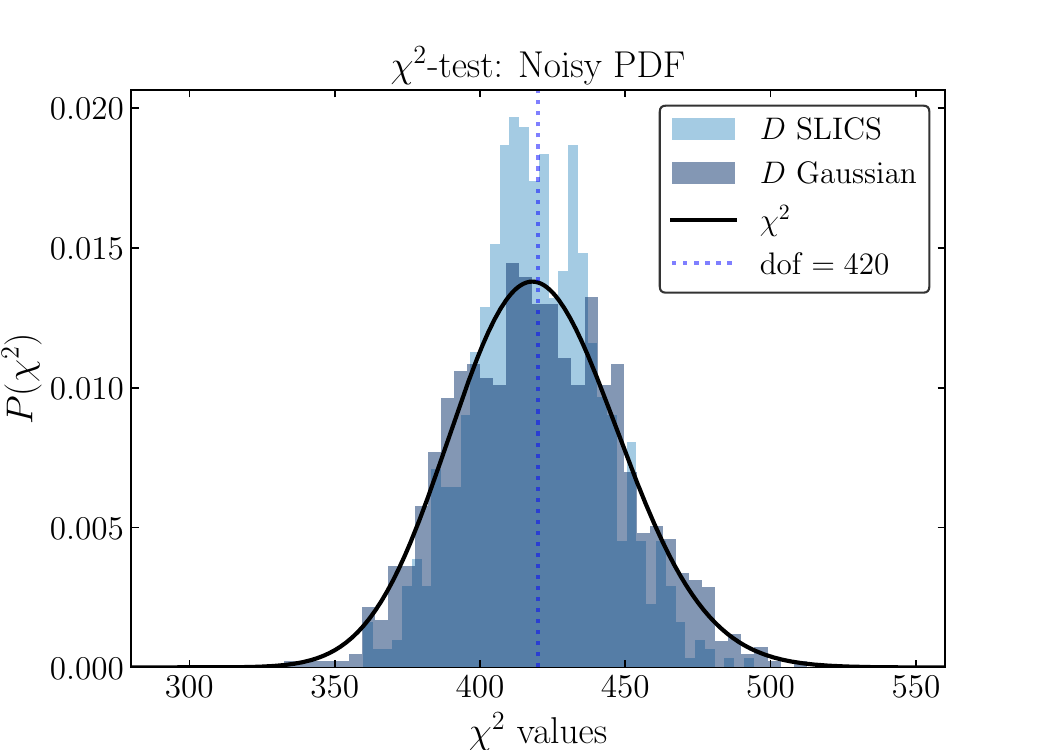}  
    \caption{$\chi^2$ test for the inverse covariance. $\chi^2-$test for the noisy $\kappa-$PDF data vector (Light blue). $\chi^2-$test for the data vector build from a multivariate normal distribution with same mean and covariance as our data vector $D$ from the SLICS measurements (dark blue). The data vector, $D$, includes three smoothing scales and 14 redshift combinations with 10 PDF bins each}.
    \label{fig:chi2_test}
\end{figure}

    In figure~\ref{fig:Gaussian_test} we demonstrate the Gaussianity of the distribution of the lowest (solid) and highest (dashed) used bins of the noisy $\kappa-$PDF for three different smoothing scales for the low redshift bin2. For each SLICS realisation we compute the fluctuation of the respective bin with respect to the mean and we normalise them with the measured variance. We then make histograms from the fluctuations and observe that they closely follow a Gaussian distribution. We perform the test for other redshift bin combinations and PDF bins obtaining similar results. For our combined tomographic $\kappa-$PDF data, we perform a $\chi^2-$test to verify the Gaussianity along with the quality of the inverse data covariance \citep[proposed in][]{Friedrich2018precisionmatrix}. If the inverse data covariance matrix is accurate enough, it should follow a $\chi^2$ distribution, which is demonstrated in figure~\ref{fig:chi2_test}. For comparison, we draw a reference sample from a data vector build from a Gaussian distribution with the same mean and covariance from SLICS measurements. Both data vectors ($D$ and $D$ Gaussian)  deviate slightly from the $\chi$-squared distribution due to sampling noise. In the Fisher forecast, we mitigate this by rescaling the inverse data covariance matrix following the Sellentin-Heavens prescription (see equation \ref{eq:Sellentin-Heavens}).

\section{$\gamma$-2PCF derivatives and  forecast}\label{app:xi}
\begin{figure}
    \centering
    \includegraphics[width=0.47\textwidth]{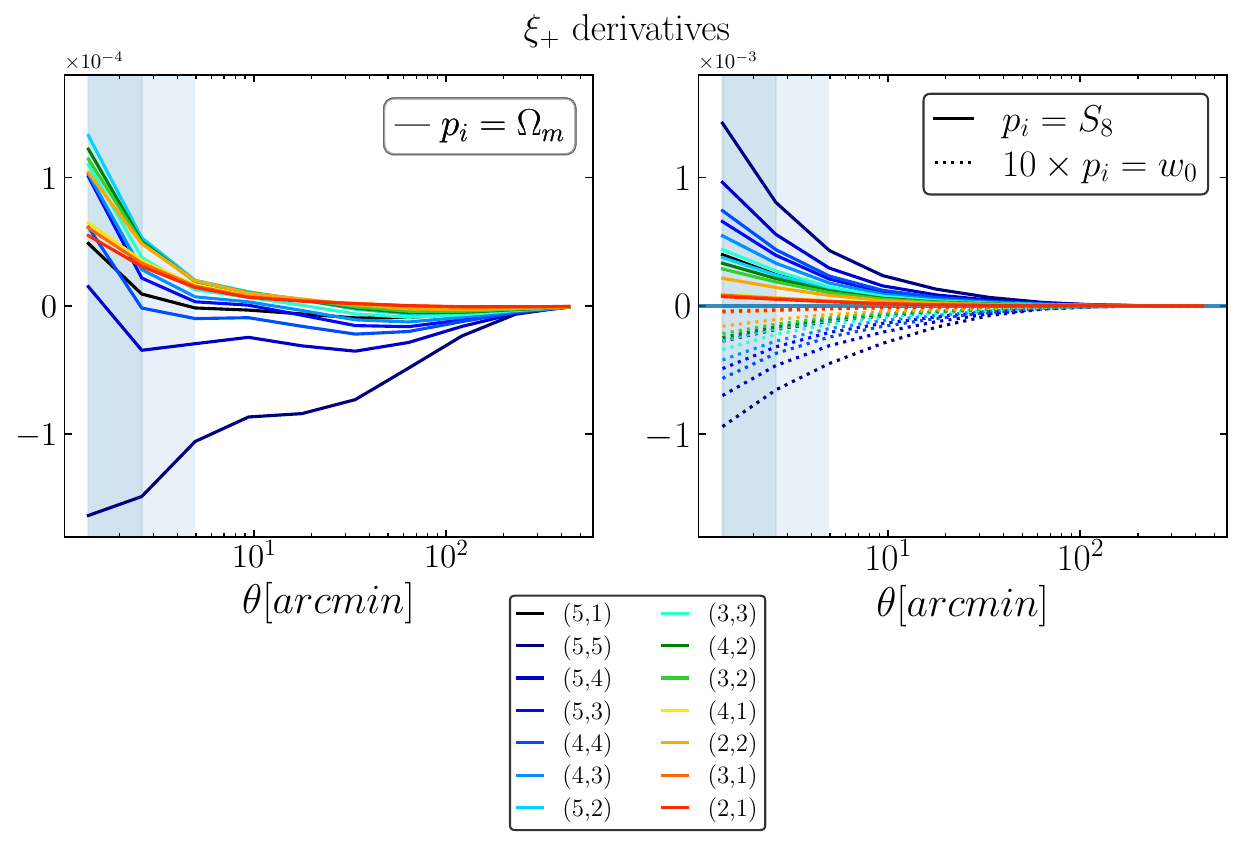}
    \includegraphics[width=0.47\textwidth]{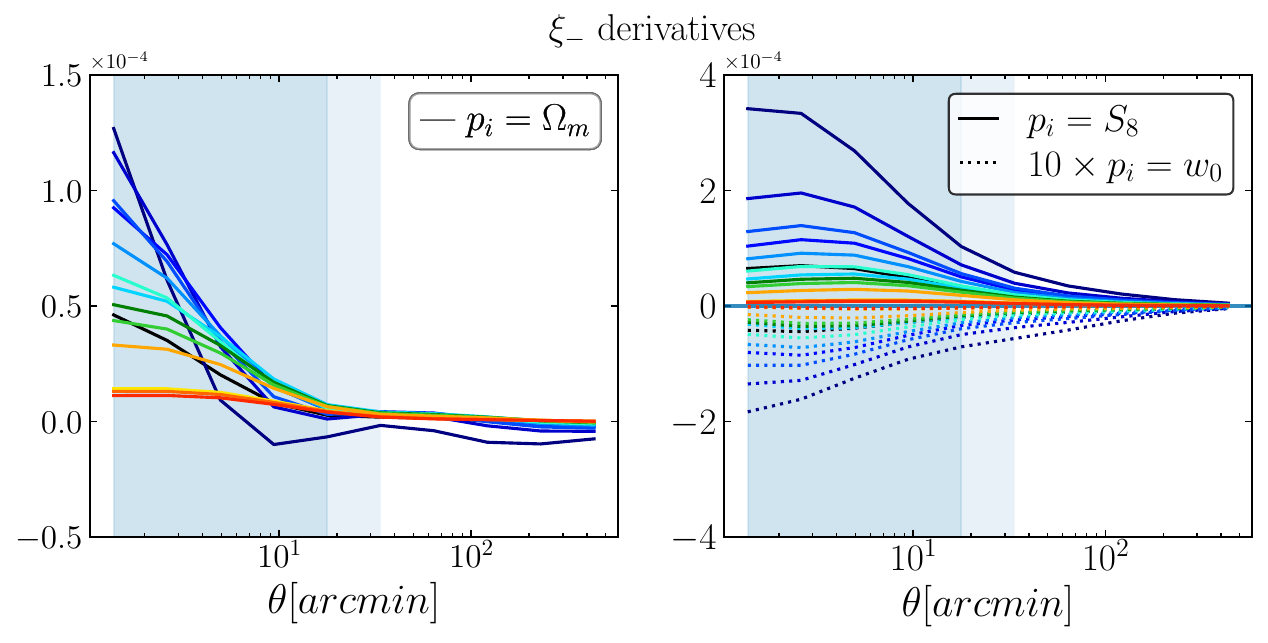}
    \caption{$\gamma-$2PCF derivatives. Upper panel: $\xi_+$ derivatives. Bottom panel: $\xi_-$ derivatives. The left panel show the derivatives with respect to $\Omega_{\rm m}$ and the right panel show the derivatives for $S_8$ and $w_0$. Note that $w_0$ has been amplified by a factor of 10 for better visualization. }
   \label{fig:xi_derivatives}
\end{figure}

\begin{figure}[h]
    \includegraphics[width=0.45\textwidth]{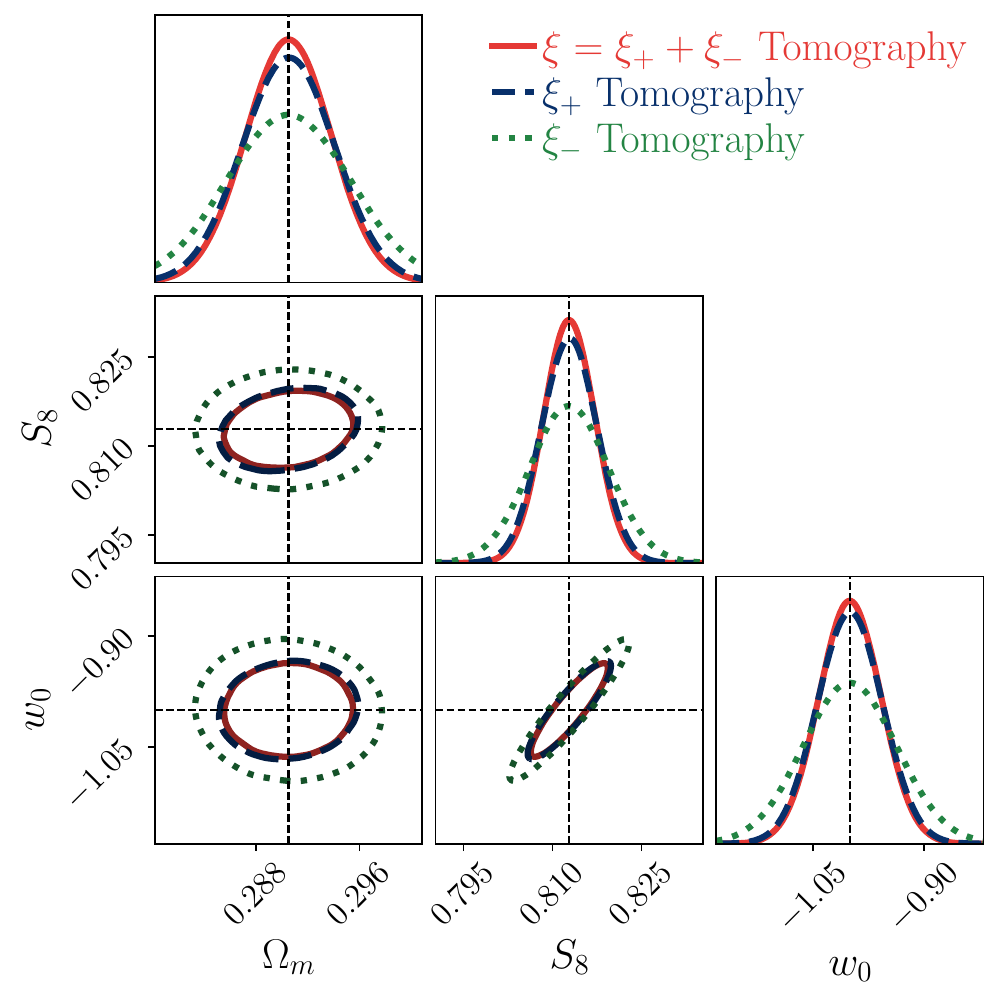} 
    \caption{$1-\sigma$ Fisher forecast constraints from the $\gamma-$2PCF tomography. Constraints from $\xi_{+}$ along (dark blue contours) for scales $\theta> ~5 $ arcmin,  constraints from $\xi_{-}$ along (dark green) for scales $\theta>~30$ arcmin. Combined constraints $\xi$ (red contours).} 
    \label{fig:xi_fisher}
\end{figure}

In figure \ref{fig:xi_derivatives} we show the $\xi_{\pm}$ derivatives for the set of parameters $p=\{\Omega_{\rm m},S_8,w_0\}$. Note that in this case we rescale the $w_0$ derivatives by a factor of $10$ for better visualisation.   
In figure \ref{fig:xi_fisher} we present the $1-\sigma$ fisher contours for $\xi_{\pm}$ and their combination $\xi$ when adding tomography. Here we use scales $\theta> ~5 $ arcmin and $\theta>~30$ arcmin (scales larger than the scales in the darker shaded region in figure \ref{fig:xi_derivatives}) for $\xi_{+}$ and $\xi_{-}$ respectively. Note that in this analysis, we are including smaller scales that were neglected in the combined analysis with the $\kappa$-PDF.  Uncertainties given by using the $\xi_{+}$ and $\xi_{-}$ statistics separately are presented in table \ref{tab:xi_fisher}. The constraining power of the $\gamma-$2PCF is dominated by the constraints from $\xi_{+}$. Combining both components of $\gamma-$2PCF provides an enhancement of the constraints of $~7\%$ with respect to $\xi_{+}$. By neglecting the smaller scales in our analysis in section \ref{sec:Fisher} and the comparison with the $\kappa-$PDF we are loosing most of the constraining power of the $\gamma-$2PCF. By adding the smaller scales, there is an improvement in the constraints given by the $\gamma-$2PCF of $~12\%$.    
\newline
\begin{table}[h]
\centering
\begin{tabular}{cccc}
\toprule 
 & $\Delta\Omega_{m}\left(10^{-3}\right)$ & $\Delta S_{8}\left(10^{-3}\right)$ & $\Delta w_{0}\left(10^{-2}\right)$\tabularnewline
\midrule
\midrule 
$\xi_{+}$ & $3.52$ & $4.60$ & $4.33$\tabularnewline
\midrule 
$\xi_{-}$ & $4.73$ & $6.59$ & $6.25$\tabularnewline
\midrule 
$\xi$ & 3.27 & $4.26$ & $4.14$\tabularnewline
\bottomrule
\end{tabular}

\caption{\label{tab:xi_fisher} Fisher 1-$\sigma$ uncertainties from the $\gamma-$2PCF tomography.}
\end{table}

\section{Additional joint $\gamma$-2PCF and $\kappa-$PDF forecast}
\label{app:joint}
\begin{figure}[h]
\centering
    \includegraphics[width=0.48\textwidth]{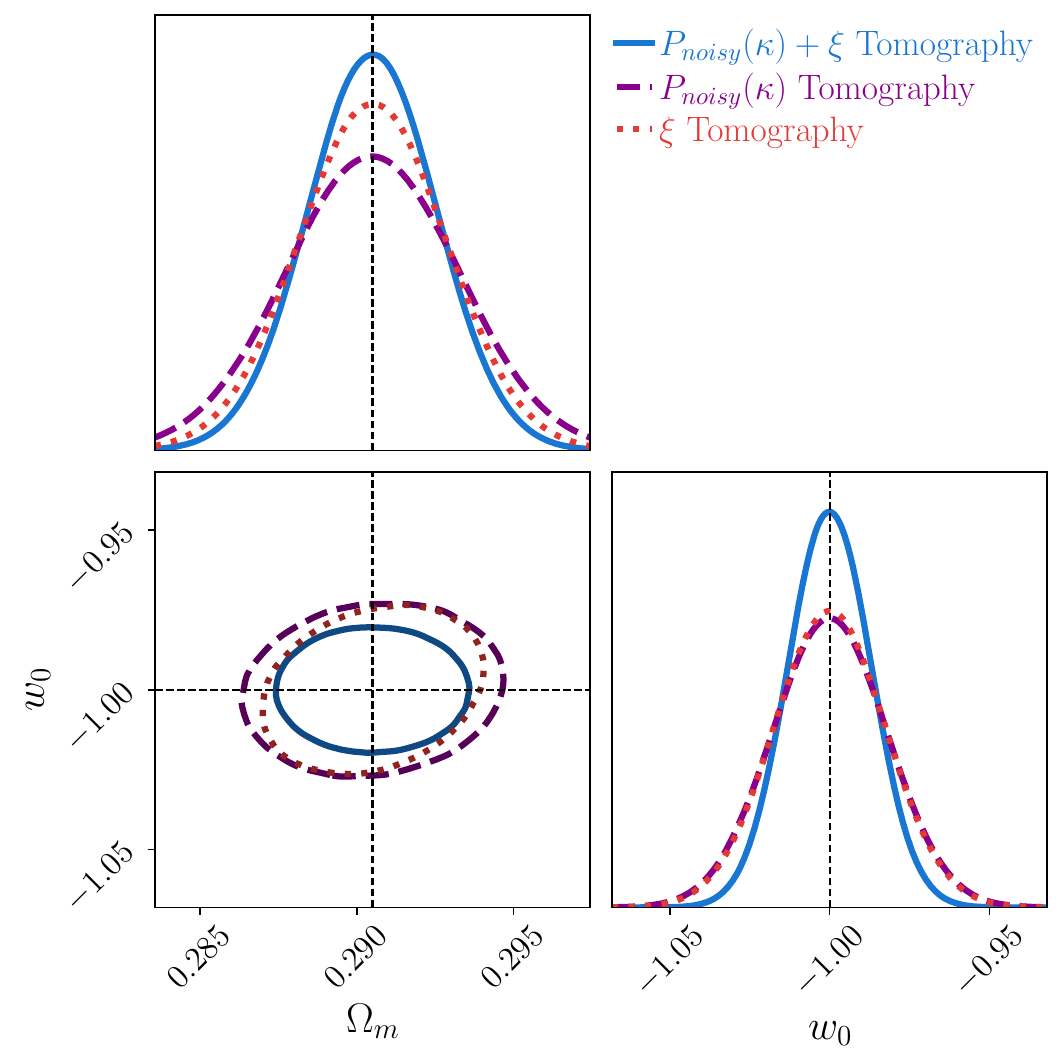} 
    \includegraphics[width=0.48\textwidth]{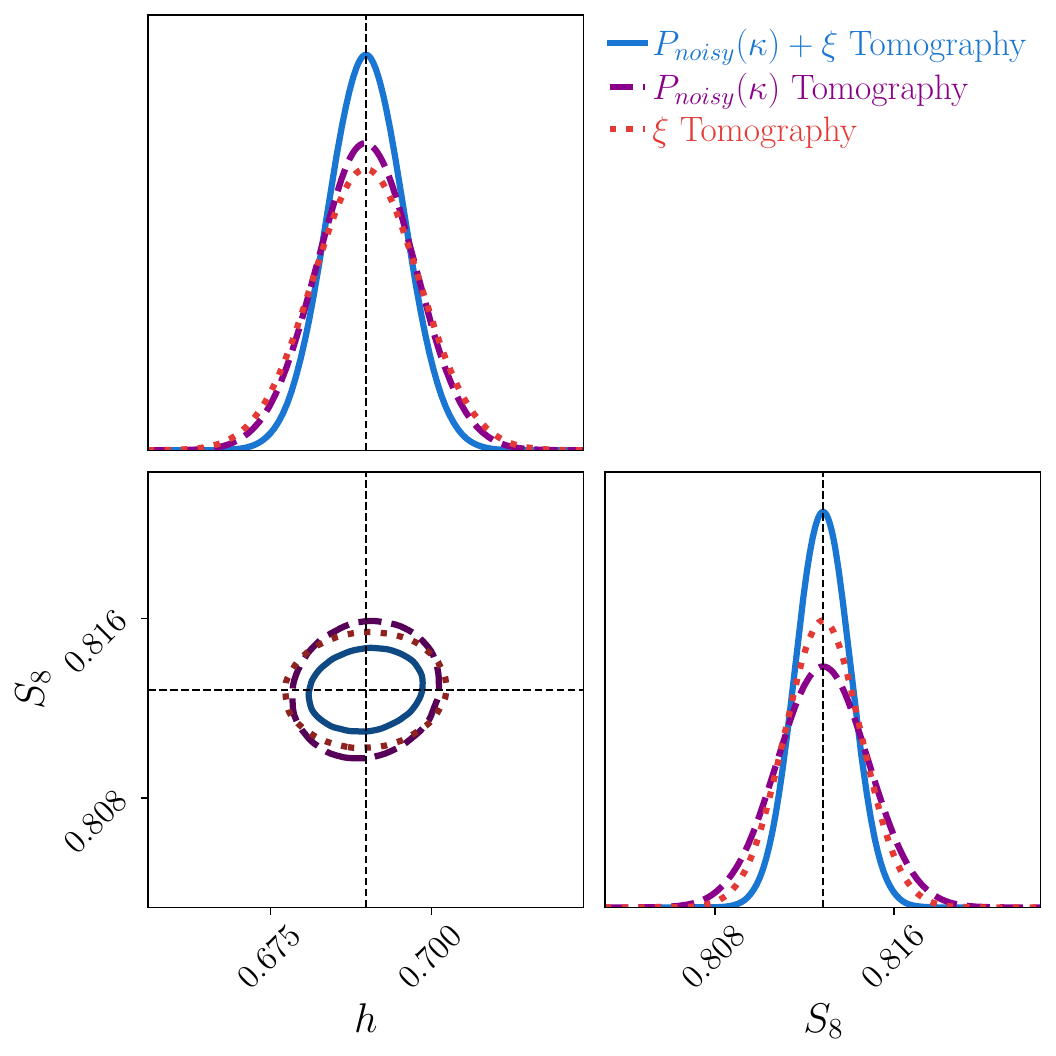}
    \caption{Fisher forecast for $1-\sigma$ constraints from the joint $\gamma-$2PCF and $\kappa$-PDF tomography as in figure~\ref{fig:fisher_joint_pdf_2pcf} but for fixed $\{S_8,h\}$ (right pannel), and $\{\Omega_{\rm m},w_0\}$ (left pannel) } 
    \label{fig:fisher_joint_pdf_2pcf_nox}
\end{figure}

Here we present extra joint $\gamma$-2PCF and $\kappa-$PDF Fisher contours.  Figure \ref{fig:fisher_joint_pdf_2pcf_nox} shows the extra cases when fixing a parameter at the time, i.e. for fixed $\{S_8,h\}$ and $\{\Omega_{\rm m},w_0\}$. The $\kappa-$PDF and the $\gamma-$2PCF give similar constraints. The $\kappa-$PDF has a slightly extra power in constraining $S_8$ and $h$ while the $\gamma-$2PCF performs slightly better in constraining $\Omega_{\rm m}$, and both provide similar constraints for $w_0$, although this similarity is broken due to the degeneracy between $w_0$ and $S_8$. As we explain in section \ref{sec:results} the $\gamma-$2PCF loses constraining power when forecasting more parameters while the $\kappa-$PDF is able to hold the constraint power outperforming the $\gamma-$2PCF.

Finally, we investigate the impact of employing a single smoothing scale for $\kappa-$PDF. In section \ref{sec:results} we combine the $\kappa-$PDF for three smoothing scales: $\{10,15,20\}$ arcmin. As the smoothing scale increases, the smoothed field becomes more Gaussian. Consequently, the $\kappa-$PDF at $10$ arcmin retains more non-Gaussian information compared to larger scales. This translates to a degradation of constraints on cosmological parameters derived from the Fisher analysis at larger smoothing scales. However, the scale dependence of derivatives with respect to $\Omega_{\rm m}$ (and $h$) introduces slight variations in the degeneracy between $\Omega_{\rm m}$ and $S_8$ and $w_0$ in the Fisher contours at different scales. This allows to partially break these degeneracies and achieve tighter constraints by combining information from multiple smoothing scales.

\begin{figure}[h]
    \includegraphics[width=0.43\textwidth]{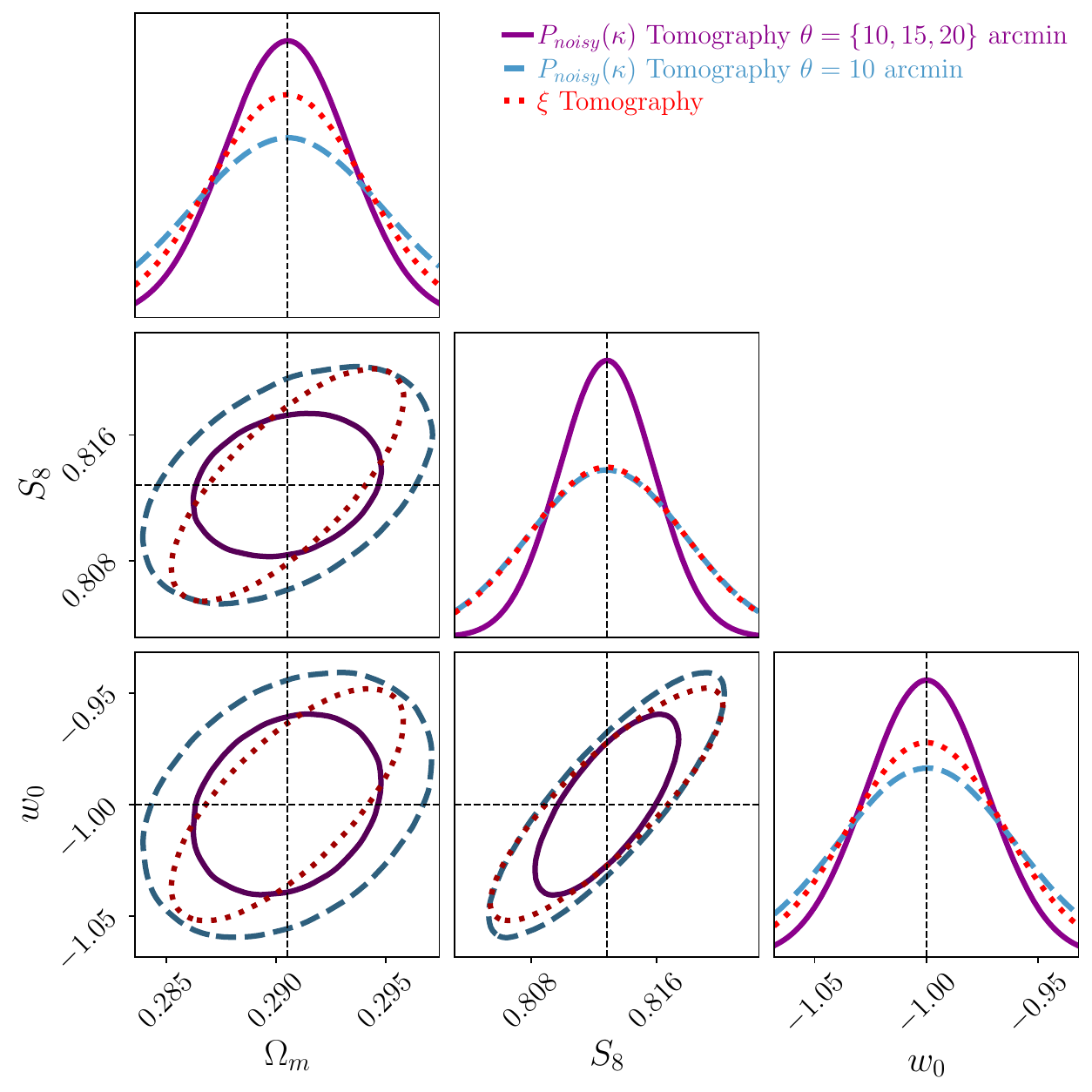} 
    \caption{Fisher forecast for $1-\sigma$ constraints from the $\kappa$-PDF considering a single smoothing scale $\theta=10$ arcmin (blue contour), and the $\kappa$-PDF combining three smoothing scales $\{10,15,20\}$ arcmin (violet contour) and the $\gamma-$2PCF and as in figure~\ref{fig:fisher_joint_pdf_2pcf} (red contour).} 
    \label{fig:fisher_joint_one_scale}
\end{figure}

Figure \ref{fig:fisher_joint_one_scale} compares the $1-\sigma$ Fisher contours for $\{\Omega_{\rm m},S_8,w_0\}$ using the $\kappa-$PDF with a single smoothing scale $\theta=10.0$ arcmin, the $\kappa-$PDF with three combined scales $\{10,15,20\}$ and the $\gamma-$2PCF. When employing the $\kappa-$PDF at a single smoothing scale for forecasting three parameters, both statistics offer comparable constraints and become complementary, leading to improved overall constraints when combined. However, if an additional parameter is included in the forecast, as mentioned earlier, the  $\gamma-$2PCF loses constraining power such that the $\kappa-$PDF at a single smoothing scale outperforms the $\gamma-$2PCF. Combining the information from three smoothing scales tighter the constraints by a factor of $\{1.5,1.6,1.4\}$ for $\{\Omega_{\rm m},S_8,w_0\}$ respectively with respect to the single scale. 

\section{Intrinsic alignment effects on the PDF}
\label{app:IA}
Besides the intrinsic shape of the source galaxies (shape noise), intrinsic alignment (IA) is also a major source of systematic noise in the shear signal. Intrinsic alignment arises from external gravitational tidal fields (galaxies are influenced by their local environment) that generate physical correlations between the shapes of neighbouring galaxies. The physics of this effect is not fully understood, and it is galaxy-type dependent, making it a significant challenge to interpret the shear signal. There are multiple models to describe the IA, as for example, the linear and non-linear alignment models \cite[LA/NLA,][]{Hirata:2007np,Bridle:2007ft}, or the  Tidal Alignment and Tidal Torquing model \citep[TATT,][]{Blazek:2017wbz}. A summary of  current IA models can be found in \cite{Lamman:2023hsj}.

Here, we study the impact of IA as described by the non-linear alignment model (NLA) on the $\kappa-$PDF. We include the effect of IA on the theoretical predictions following the prescription given in section 6.1 of \cite{Barthelemy:2023mer} and we validate the model with the measurements from a set of \textit{cosmo}-SLICS that has been infused with IA signal. 

The NLA model assumes that galaxies preferably align with the surrounding tidal fields in a redshift dependent way, such that the contributions of the IA to the ellipticity are given by \cite{Harnois-Deraps:2021krd}
\begin{align}
    \epsilon_{\rm IA,1} & = A(z)(s_{11}-s_{22})\,, \\\qquad \epsilon_{\rm IA,2} & = 2A(z)s_{12}\,,
\end{align}
where $s_{ij} = \de_i \de_j \Phi (\bm{x}) $ are the Cartesian components of the tidal tensor. $A(z)$ encodes the redshift-dependent scaling as
\begin{equation}
\label{eq:Az}
    A(z) = -A_{\rm IA} \left(\frac{1+z}{1+z_0}\right)^{\alpha_{\rm IA}} \frac{\bar{C}  \bar{\rho}(z)}{D(z)}\,,
\end{equation}
with $A_{\rm IA}$ the strength of the tidal coupling and $\alpha_{\rm IA}$ that describes the redshift evolution, $z_0$ is a pivot redshift, $\bar{\rho}(z)$ is the mean density at redshift $z$  and $\bar{C}$ is a calibrated constant chosen for historical reasons \citep[see][]{Brown:2000gt}. This formalism is valid in the linear regime, this assumption extends to the non-linear case by preserving the functional form of the effect and using a non-linear potential.

The infusion of the IA signal into the simulation catalogues relies on the computation of the trace-free projected tidal field. The projected tidal field maps are constructed for each of the mass sheets in the light-cones of the simulation and interpolated at each galaxy position, such that the observed ellipticity is given by
\begin{align}
{\boldsymbol \epsilon}_{\rm obs} = \frac{\boldsymbol{ \gamma} + \boldsymbol{ \epsilon}_{\rm int}}{1 + \boldsymbol{\gamma}^* \boldsymbol{\epsilon}_{int}} \, ,
\end{align}
where $\boldsymbol{ \epsilon}_{\rm int}$ is the intrinsic galaxy ellipticity composed by the shape noise and the IA signal as
\begin{equation}
    \boldsymbol{ \epsilon}_{\rm int} = \frac{\boldsymbol{\epsilon}_{\rm IA} + \boldsymbol{\epsilon}}{1+\boldsymbol{\epsilon}_{\rm IA}\boldsymbol{\epsilon} }\,,
\end{equation}
we choose $A_{\rm IA}=1$ and $\alpha_{\rm IA} = 0$.

In the NLA model, as the tidal fields are related to the matter density contrast, the IA field $\delta_{\rm IA}$ can be written as a bias effective expansion as \cite{}
\begin{equation}
    \delta_{\rm IA}(\chi \bm{\theta},\chi) = A(z) \delta(\chi \bm{\theta},\chi)\,,
\end{equation}
where the bias factor is given in equation \eqref{eq:Az}. The contribution to the convergence field is then written as
\begin{equation}
    \kappa_{\rm IA}(\chi \bm{\theta},\chi) = \int^{\infty}_{0} \,dz\frac{d\chi}{dz}\, n(z) \delta_{\rm IA}\left(\chi \bm{\theta},\chi\right)\,,
\end{equation}
such that we can include the effect of IA in the predicted PDF by modifying the projection kernel in equation \eqref{eq:weight_function} as 
\begin{equation}
    w(\chi) \rightarrow{} w(\chi) + A(z) n(z)\,.
\end{equation}
We now just re-use the framework described in section \ref{sec:kappa_PDF_th} and Appendix \ref{app:CGF} to predict the IA-infused $\kappa-$PDF with our modified lensing kernel.

The proposed method to include the NLA model at the PDF level allows to consistently infuse the IA effect on the full hierarchy of the high-order cumulants. Nevertheless, it is interesting to note that this is approximately equivalent to a modification of only the standard deviation in the CGF, although with a different scaling than what we traditionally expect from tree-order perturbation theory and equivalently our LDT formalism. Indeed we find that no matter the sources distribution, the PDF of $\kappa/\sigma$ is mostly unchanged by the inclusion of the IA effect. This can be seen by writing 
\begin{equation}
    \frac{\kappa}{\sigma} = \frac{\int {\rm d}\chi \, (\omega(\chi)+ A(z(\chi)) \cdot n(z(\chi))) \delta}{\sqrt{\int {\rm d}\chi \, (\omega(\chi)+ A(z(\chi))^2 \cdot n(z(\chi))) \langle\delta^2\rangle_c}}.
\end{equation}
\newline
From there and given the `peaked' nature of the lensing kernels with or without IA, it is possible to approximate both integrals in the numerator and denominator by their integrand values at the peak of the kernels. This leads to $\frac{\kappa}{\sigma} \approx \frac{\delta}{\sqrt{\langle\delta^2\rangle_c}}|_{\chi \, \text{where} \, \omega'(\chi) = 0}$. Finally, given that the lensing kernel peaks roughly halfway between the observer and the mean source, the NLA correction typically does not change the peak location so that this type of modification of the kernel mostly leaves $\kappa/\sigma$ unchanged.

\begin{figure}
    \includegraphics[width=0.45\textwidth]{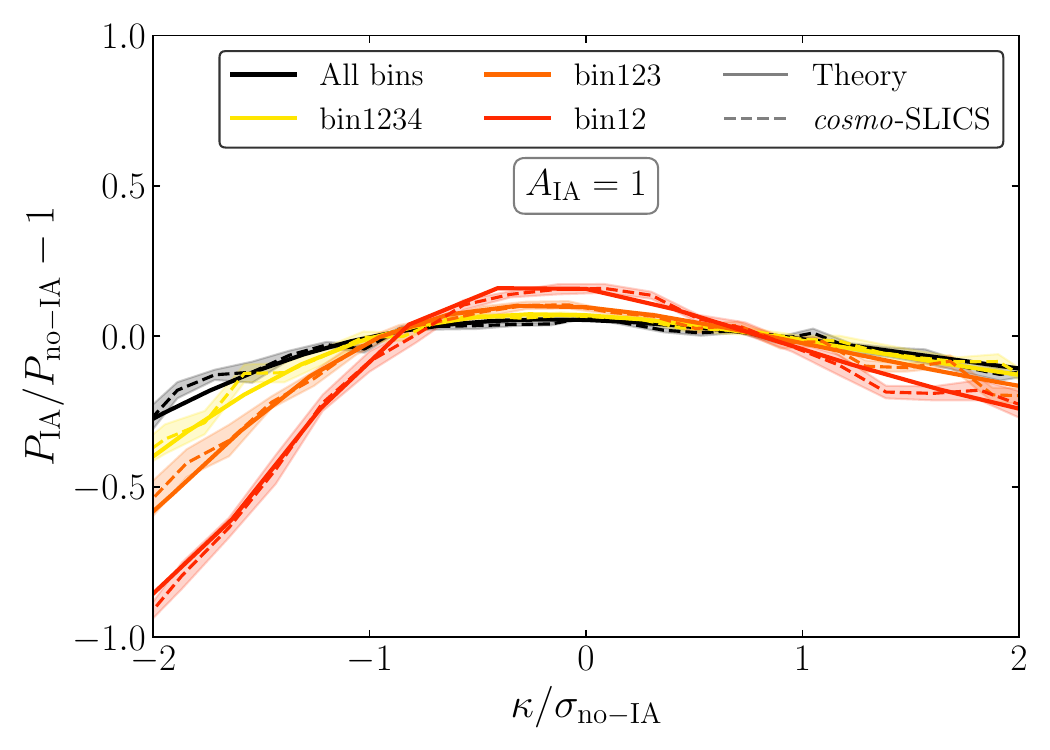} 
    \caption{Effect of IA on the $\kappa-$PDF. (Dashed lines) Residual between the $\kappa-$ PDF measured from \textit{cosmo}-SLICS simulations with and without infused IA. All measurements are normalized by the variance of the measure PDF without IA. (Solid lines) Residuals between the NLA model in the predicted $\kappa-$PDF and theoretical predicted $\kappa-$PDF without IA, normalized by the variance of the theory PDF without IA. We show results for the redshift combinations that include bin1, which is the most affected by IA. The shaded area represents the standard deviation of the mean across realizations.}
    \label{fig:ratio_IA_theory}
\end{figure}

\begin{figure}
      \centering
    \includegraphics[width=\columnwidth]{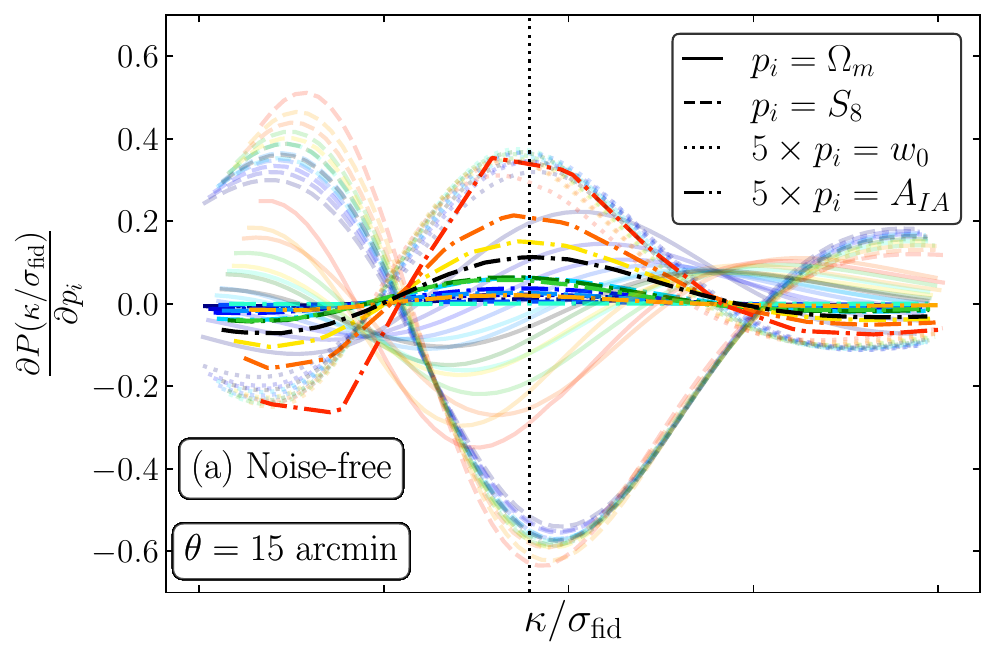}
      \caption{Comparison of $\kappa-$PDF derivatives w.r.t. cosmological parameters $\Omega_m$ (solid), $S_8$ (dashed), $w_0$ (dotted lines) and the strength of intrinsic alignment $A_{IA}$ (dashdoted line) for all  tomographic bins where the colour map .}
      \label{fig:derivative_IA}
  \end{figure}

\begin{figure}
        \centering
       \includegraphics[width=0.45\textwidth]{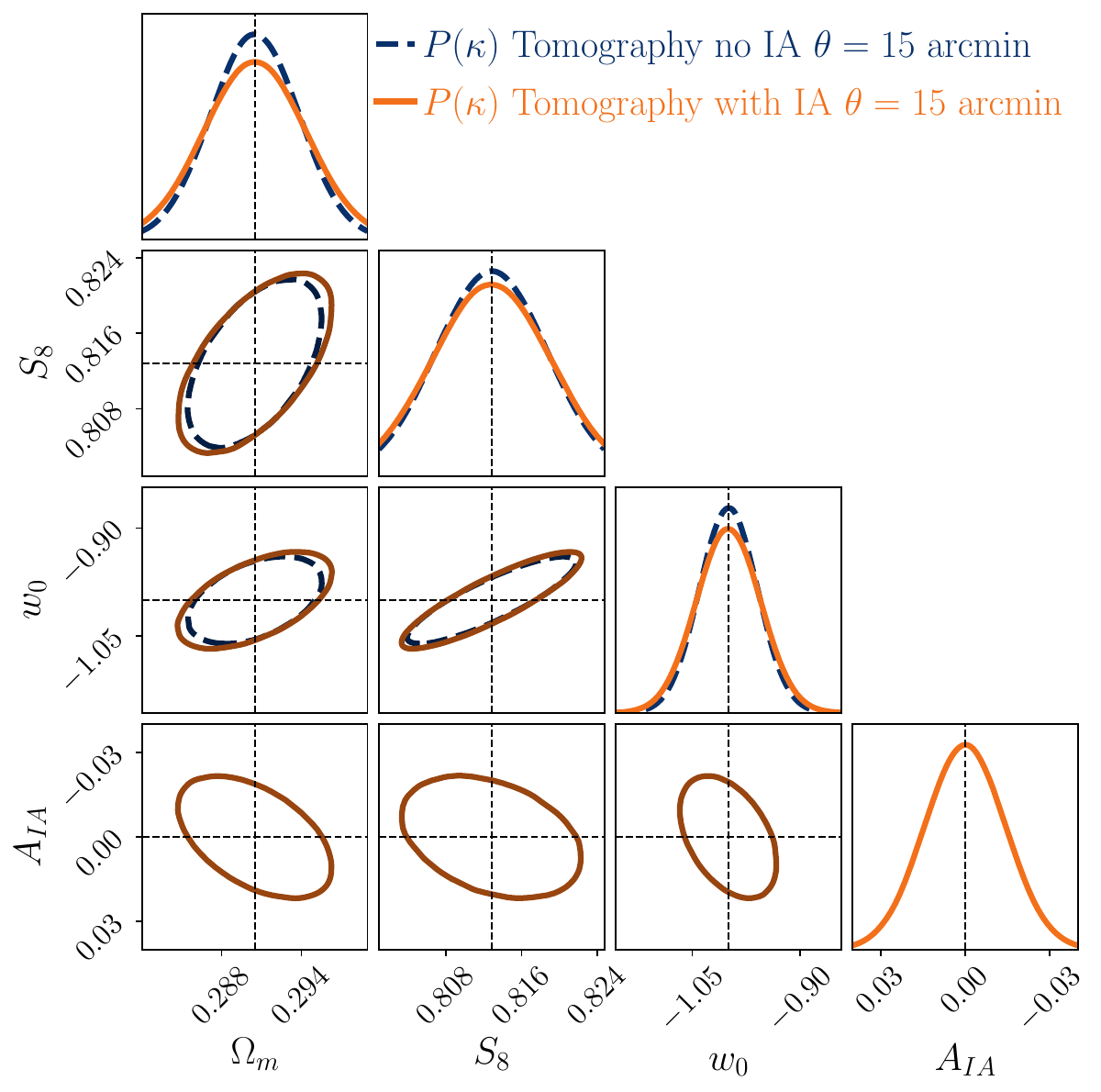}
        \caption{Fisher forecast for $1-\sigma$ constraints from the $\kappa-$PDF including IA at $\theta = 15$ arcmin. Constraints on the cosmological parameters $\{\Omega_m,S_8,w_0,A_{IA}\}$, without the IA signal (blue contours) and including the IA signal (orange contours).}
        \label{fig:IA_Fisher_AIA}
\end{figure}

Figure \ref{fig:ratio_IA_theory} shows the residual between the $\kappa-$PDF  with and without infused IA, for both the NLA theory predictions and the \textit{cosmo}-SLIC measured PDF averaged over 10 realizations (we exclude shape noise). As predicted by the NLA regime, the lowest redshift bin (bin1) shows the largest impact, affecting all combinations that include it. Hence, we only show results for the redshift combinations that include bin1. The proposed model to include IA in the $\kappa-$PDF predicts the measured $\kappa-$PDF with infused IA with high accuracy.
\newline
To gain an intuition on the IA impact on the  cosmological parameter constraints, we perform a Fisher forecast including $A_{IA}$. We employ noise-free $\kappa-$PDF prediction for the cosmo-SLICS cosmology with and without IA signal. We compute derivatives with respect to $A_{IA}$ from the $\kappa-$PDF including the IA signal for $A_{IA}=1.0$ and the fiducial $A_{IA}=0.0$. Because the impact of $A_{IA}$ is driven by the contribution of the first tomographic bin, it leads to derivatives with a distinct dependence on the tomographic bin as shown in Figure~\ref{fig:derivative_IA}. In Figure \ref{fig:IA_Fisher_AIA} we show the resulting Fisher contours including $A_{IA}$. The tomographic convergence PDF provides quite tight constraints on $A_{IA}$. After marginalising over the IA parameter the constraints on $\{\Omega_m,S_8,w_0\}$ are degraded slightly by  a factor of $\{1.15, 1.07, 1.11\}$, respectively, and we observe barely any effect on the direction of the ellipses. Although the effect on the cosmological parameter constraints is small ignoring the impact of IA could lead to biases, therefore it is necessary to include it into the modelling and investigate further to test our expectations.
However, note that this forecast does not include shape noise and additional systematics such as photo-z errors which may be largely degenerate with the IA signal and hence degrade constraints more significantly. Additionally, we are not taking into account the large projection effect when considering the redshift dependence of the IA amplitude.

\bibliography{main.bib}
\end{document}